\documentclass[preprint]{aastex}

\shortauthors{Skillman, C\^ot\'e, \& Miller}
\shorttitle{Star Formation in Sculptor Group Dwarfs}

\begin{document}

\title{Star Formation in Sculptor Group Dwarf Irregular Galaxies and
the Nature of ``Transition'' Galaxies}

\author{Evan D. Skillman}
\affil{Astronomy Department, University of Minnesota,
    Minneapolis, MN 55455} 
\email{skillman@astro.umn.edu}

\author{St\'ephanie C\^ot\'e}
\affil{Canadian Gemini Office, HIA/NRC of Canada,
5071 West Saanich Rd., Victoria, B.C., Canada, V9E 2E7}
\email{Stephanie.Cote@hia.nrc.ca}

\and

\author{Bryan W. Miller}
\affil{AURA/Gemini Observatory, Casilla 603, La Serena, Chile;}
\email{bmiller@gemini.edu}

\begin{abstract}

We present new H$\alpha$ narrow band imaging of the HII regions in eight
Sculptor Group dwarf irregular (dI) galaxies.  The H$\alpha$ luminosities
of the detected HII regions range from some of the faintest detected 
in extragalactic HII regions ($\sim$ 10$^{35}$ erg s$^{-1}$ in SC~24) to
some of the most luminous ($\sim$ 10$^{40}$ erg s$^{-1}$ in NGC~625).
The total H$\alpha$ luminosities are converted into current star formation
rates (SFR).  Comparing the Sculptor Group dIs to the Local Group dIs, we
find that the Sculptor Group dIs have, on average, lower values of
SFR when normalized to either galaxy luminosity or gas mass (although 
there is considerable overlap between the two samples).
The range for both the Sculptor Group and Local Group samples
is large when compared to that seen for the sample of gas-rich quiescent
low surface brightness (LSB) dIs from van Zee et al.\ (1997) and the sample of
isolated dIs from van Zee (2000, 2001).  This is probably best understood
as a selection effect since the nearby group samples have a much larger
fraction of extremely low luminosity galaxies   
and the smaller galaxies are much more liable to large relative
variations in current SFR. 
The Sculptor Group and LSB samples are very similar with regard to
mean values of both $\tau_{gas}$ and $\tau_{form}$, and 
the Local Group and isolated dI samples are also similar to each other
in these two quantities.
Presently, the Sculptor Group lacks dI galaxies with elevated normalized
current SFRs as high as the Local Group dIs IC~10 and GR~8.

The properties of ``transition'' (dSph/dIrr) galaxies in Sculptor and the Local
Group are also compared and found to be similar.   The transition galaxies
are typically among the lowest luminosities of the gas rich dwarf 
galaxies.  Relative to the dwarf irregular galaxies, the transition 
galaxies are found preferentially nearer to spiral galaxies, and are 
found nearer to the center of the mass distribution in the local cloud. 
While most of these systems are consistent with normal dI galaxies which 
currently exhibit temporarily interrupted star formation, the observed
density-morphology relationship (which is weaker than that observed
for the dwarf spheroidal galaxies) indicates that environmental
processes such as ``tidal stirring'' may play a role in causing their 
lower SFRs.

\end{abstract}


\section{Introduction}

Low-mass dwarf irregular galaxies provide an important testing ground
for several fundamental questions about star formation, star formation
rates (SFRs), galactic evolution, and cosmology.  
Due, in large part, to attempts to understand the possible evolutionary
connections between the dwarfs with negligible or extremely low present
SFRs (the dSph and dE galaxies, hereafter dE galaxies)
and the dwarfs with obvious signs of present star formation (the dIrrs,
blue compact dwarfs, HII galaxies, hereafter dI galaxies), many theorists 
are turning their
attention to the problem of dwarf galaxy evolution.  Environmental effects
are turning out to be a key parameter (e.g., van den Bergh 1994b; 
Klypin et al.\ 1999; Moore et al.\ 1999; Gnedin 2000; Mayer et al.\ 2001a,b; 
Carraro et al.\ 2001).
An especially interesting question in this regard is the effect of
reionization on the suppression of dwarf galaxy formation (Efstathiou
1992; Babul \& Rees 1992; Quinn, Katz, \& Efstathiou 1996; 
Barkana \& Loeb 1999; Bullock et al.\ 2000).
By comparing the properties of dwarf galaxies in different environments
we may be able to isolate key environmental variables (e.g., local density,
companionship, group vs.\ cluster membership) in order to
constrain these theories.

The Sculptor group is the closest group of galaxies beyond our Local Group.
Originally thought to be at a distance of 2.5 Mpc (de Vaucouleurs 1975),
the main members of the group are 5 late-type spiral galaxies (NGC~55, 
NGC~247, NGC~253, NGC~300, and NGC~7793) with distances ranging between 
1.6 and 3.4 Mpc (Puche \& Carignan 1988 and references therein) and
a dwarf spiral (NGC~45) at a distance of 4.4 Mpc.  Additionally, NGC~24,
another dwarf spiral, lies close to NGC~45 in radial velocity and projected 
location and is usually included as a group member (e.g., Giuricin et al.\
2000) although some distance estimates put it as far away as 11 Mpc
(Puche \& Carignan 1988).  Jerjen, Freeman, \& Binggeli (1998) studied 
surface brightness fluctuations in 5 
Sculptor group dEs and determined a spread in distances of dE
members of the group from 1.7 to 4.4 Mpc (but note that Jerjen \& 
Rejkuba 2001 claim that some of these distances should be revised).  
A well defined relationship
between radial velocity (corrected to the galactocentric standard of rest)
and distance was discovered for all members of the Sculptor group (Jerjen
et al.\ 1998), indicating that radial velocity may be a good distance
indicator for Sculptor Group members.  Jerjen et al.\ (1998) propose that
the Sculptor Group and the Local Group are part of the same Supergalactic
structure (a part of the Coma - Sculptor Cloud delineated by Tully \&
Fisher 1987).  A dynamical picture of the Sculptor Group can be found in 
Whiting (1999).

Miller (1994), C\^ot\'e (1995), and Jerjen et al.\ (2000) independently 
searched for and found new dwarf galaxy members of the Sculptor Group.  
The Sculptor Group is now known to contain 16 dwarf irregular galaxies 
(C\^ot\'e et al.\ 1997), some of which are amongst the lowest luminosity  
dwarf irregulars known.   The present Sculptor Group
membership situation can be found in Jerjen et al.\ (1998; 2000).

Miller (1996) noted that several Sculptor Group dIs were undetected in 
H$\alpha$ emission at reasonably sensitive levels.  This has several 
implications for the present study.  
First, as Miller (1996)  pointed out, the Sculptor
Group dIs may have relatively low average SFRs when
compared to Local Group dIs. 
Second, although some of the Miller (1996) sample have now been detected in
H$\alpha$ at low levels (see \S 2.2), there remain three dwarf galaxies 
in the Sculptor Group (SDIG, DDO~6, and UGCA 438) with the properties
of ``transition type'' galaxies (no detectable HII regions and 
M$_{HI}$/L$_B$ values in the range of dI galaxies).  Thus, the 
Sculptor Group sample allows us to investigate this less common type
of galaxy (see, e.g., Miller et al.\ 2001).
Third, H$\alpha$ surveys identifying HII 
regions are a necessary first step in studying the ISM abundances of 
galaxies.  Since relatively high surface brightness HII regions are 
usually required for an accurate abundance analysis, this means that
we may not be able to accurately measure the chemical abundances in 
all of the known Sculptor Group dIs.  
 
Here we present deep CCD H$\alpha$ imaging of eight Sculptor group
dIs obtained at the ``Danish''
1.5m telescope located at the European Southern Observatory.
We have detected HII regions in all eight dIs observed, 
in addition to the two dIs detected in H$\alpha$
by Miller (1996).  We present coordinates and fluxes for the HII regions
and estimated SFRs for these galaxies.
Additionally, we have used the CTIO 4-m to obtain optical spectra of
ten HII regions located in five of these dIs 
and these observations are presented in a companion paper. 
By comparing observations of the dwarf galaxies in the next nearest
group to other well defined samples, we hope to better understand basic 
questions such as: what
are the average properties of the dwarf galaxies? and, are there
observable signatures of environmental influences on galaxy evolution? 

\section{Observations}

\subsection{Target Selection}

All galaxies observed were chosen from the list of C\^ot\'e et al.\ (1997).
Table 1 lists some of the properties of the observed galaxies.
Based on the observation by Jerjen et al.\ (1998) that the 
scatter in the distance -- velocity relationship for the Sculptor Group
members is ``remarkably small'',
we have derived new distances for these galaxies using the
recessional velocities listed in C\^ot\'e et al.\ (1997) and the
formula given in Jerjen et al.\ (1998):
\begin{equation}
v_{\rm GSR}\ [km s^{-1}] = 119\, (\pm 7)\, D\ [Mpc]\,  - 136\, (\pm 14).
\end{equation}
For the nine galaxies (6 spirals and 3 dwarfs) used to define the relationship, 
the average difference between the measured distance and the estimated distance 
is 10\%.  
Eight of the nine galaxies show differences of less than 20\% (the exception 
being NGC 45, with a difference of 21\%).
Dolphin (private
communication) has calculated distances from Hubble Space Telescope
observations of the red giant branch tip
for five (1 spiral and 4 dwarfs) Sculptor group galaxies with known recession
velocities (DDO~6, DDO~226, NGC~253, ESO~245-G005 and ESO~294-G010), and these 
five galaxies show good agreement with the relationship derived
by Jerjen et al.\ (1998); all five galaxy distances are within 20\% of that
predicted by the relationship.  The relationship should be updated with 
inclusion of these new data, but, based on the above, the anticipated 
revision will be small (of order the uncertainty in the 
relationship), and we will use the relationship as published.
The estimated errors in distances are small enough that we can 
begin to examine positional relationships between galaxies.
Note that the finding by Jerjen \& Rejkuba (2001) that the distances
estimated for some of the dEs via the method of surface brightness
fluctuations could be significantly in error does not directly imply that
the formula relating distance and recessional velocity needs to be
revised since most of the dEs do not have measured recession velocities
and were not used in calibrating the relationship.

This equation was derived for galaxies covering the range in corrected
recessional velocity from $\sim$ 70 to $\sim$ 490 km s$^{-1}$, and 
thus may not be strictly valid for the three galaxies observed here
with corrected recessional velocities in excess of 500 km s$^{-1}$.
The fact that several of the dIs have recessional velocities in 
excess of 500 km s$^{-1}$ implies that a significant number of dI 
galaxies lie on the far side of the Sculptor group.  
(Note that the spiral galaxy NGC 24, with a recession velocity of 
555 km s$^{-1}$ is also in this region.) 
In comparison with 
the Local Group, this is not unexpected, as several of the Local Group 
dIs lie in the  low density periphery of the Local Group at distances 
in excess of 1 Mpc from the Milky Way and M31 (Mateo 1998).
The absolute B magnitudes in Table 1 were calculated from the total B 
magnitudes and extinctions listed in C\^ot\'e (1995).  From a comparison 
of the values in C\^ot\'e (1995) and other values in the literature,
the uncertainty in M(B) from the photometry is probably a little less 
than 0.2 magnitudes, and, when considering the error in the distance
estimates, the total error is probably a little larger than 0.2 magnitudes
for the typical Sculptor Group member.  The error could be larger 
for those galaxies with velocities in excess of 500 km s$^{-1}$.
Nonetheless, these errors are small relative to the dynamic range of
the luminosities of the galaxies and allow us to 
investigate relationships between intrinsic properties of galaxies. 

\subsection{H$\alpha$ Imaging}

The search for HII regions in the Sculptor dIs was 
performed on October 24 and 25, 1995 with DFOSC at the 1.5m ``Danish" telescope 
at the European Southern Observatory.  Additional data for one galaxy 
(ESO~473-G24) were gathered on November 4, 1995, courtesy of Chris Lidman. 
The detector used was the Loral 2048 $\times$ 2048, giving a scale of 
0.39\arcsec\ pixel$^{-1}$ and field of view of 13.6\arcmin\  $\times$ 
13.6\arcmin .  The CCD was read 
out in the high gain mode yielding a readout noise of 7.2 e$^-$ r.m.s.\ with 
a gain of 1.31 e$^-$/ADU.  The galaxies were imaged through a narrowband 
H$\alpha $ filter with a FWHM of 62.1 \AA\  centered at $\sim $6561 \AA\ 
(ESO\#693), and continuum off-band images were taken in Gunn {\it i}.
The observations usually consisted of 2 $\times$ 1200s for the H$\alpha $ 
images and 2 $\times$ 600s for the {\it i}-band images. The seeing varied 
between $\sim $1\arcsec\ and 1.3\arcsec , but the conditions were not 
photometric.

The data were reduced following the usual procedures with IRAF\footnotemark. 
After registering the different images,
the two H$\alpha $ and two {\it i } frames 
were co-added. To optimize the continuum subtraction, we used a small 
kernel Gaussian filter to smooth the {\it i } image in order to match as
closely as possible the point-spread functions of the {\it i } and H$\alpha $ 
images.  Half a dozen isolated bright stars were used to determine a scaling 
factor between the {\it i } and H$\alpha $ frames.  The final continuum 
subtracted image was then produced by subtracting the appropriately 
scaled {\it i } image from the H$\alpha $ image. Figures 1a-d show the 
original H$\alpha $ and the continuum-subtracted H$\alpha $ images. 
\footnotetext{IRAF is distributed by the National Optical Astronomy 
Observatories, which are operated by the Association of Universities for 
Research in Astronomy, Inc., under cooperative agreement with the National
Science Foundation.}

Although many HII regions are fairly well isolated, many are assembled in 
larger complexes, or have more complex morphologies like loops or filaments, 
as is commonly seen in late-type galaxies.
Hence it was necessary to determine the boundaries of each HII region by 
eye, using POLYPHOT, selecting each distinct emission peak as a separate 
HII region. The borders of the HII regions were set to a constant H$\alpha $ 
surface brightness level for each galaxy but because the nights were not
photometric these levels are estimated to be varying around $5 \pm 2 \times 
10^{-17}$ erg cm$^{-2}$ s$^{-1}$ arcsec$^{-2}$. The positions and fluxes 
of the HII regions are given in Table 2.  Errors in the fluxes are estimated
to be of order 
20\% due to photometric uncertainties, except for SC~18, SC~24, and NGC~625;
these galaxies were observed
at the end of the second night when the conditions had improved, and 
the errors are estimated to be 10\%.  In fact, for NGC~625, which 
is the only galaxy in our sample with previous  H$\alpha $ measurements 
in the literature, our fluxes agree within 4\% with that of Marlowe 
et al.\ (1997).  On the other hand, for AM~0106-382 the conditions
had deteriorated so badly that the galaxy was not even detectable in the 
continuum {\it i} exposures.  However, our H$\alpha $ images were compared 
with the $i$ data obtained by C\^ot\'e (1995), and it appears that the emission 
peaks seen in our H$\alpha $ image do not have any counterparts in the $i$ 
image. This indicates that they could be genuine HII regions, so 
they are listed as well in Table 2.  None of the fluxes given in Table 2 have 
been corrected for [NII] contamination; these dwarf galaxies have typically
very low nitrogen abundances (see companion paper), so this introduces an 
additional $\sim $6\% flux uncertainty.  For deriving accurate positions 
for the HII regions, HST Guide Star Reference Frame scans from
the STScI Digitized Sky Survey\footnotemark\ were used, and were 
compared with positions 
obtained similarly by deriving astrometric plate solutions using bright 
stars in the Automatic Plate Measuring (APM) catalog (cf.\ van Zee 2000). 
In each case these positions agreed to within less than 2\arcsec .
\footnotetext{The Digitized Sky Surveys were produced at the Space Telescope
Science Institute under U.S.  Government grant NAG W-2166. The images of 
these surveys are based on photographic data obtained using the Oschin 
Schmidt Telescope on Palomar Mountain and the UK Schmidt Telescope.
The plates were processed into the present compressed digital form with
the permission of these institutions.}

The HII regions in the two faintest galaxies (SC~18 and 
SC~24) were just at the limit of detectability.  In both cases, 
coincidence with continuum sources cast doubt on whether these are
true H$\alpha$ sources or possible artifacts of an imperfect
continuum subtraction.  Further inspection and experimentation
led to confidence that these are indeed real H$\alpha$ sources.
Nonetheless, spectroscopic observations are required for confirmation.  
For the present paper, we will treat these H$\alpha$ sources as
bona fide HII regions.  Note that in the case of the Local Group dI
DDO~210, van Zee et al.\ (1997) found a similar single H$\alpha$
detection to show broad Balmer lines and no forbidden lines,
and they speculate that this source may be a luminous blue variable star.
  
Miller (1996) observed and detected HII regions in the Sculptor group dIs
UGCA 442 ($=$ ESO~471-G06) and ESO~245-G05 ($=$ A 143), but did not
detect H$\alpha$ emission from six other Sculptor group dIs.
Note that we did {\it not} re-observe any of the  Sculptor group dIs 
from Miller's sample.  Miller's non-detection of HII regions in
the Sculptor Dwarf Irregular Galaxy ($=$ SDIG, ESO~349-G31) has been
confirmed by Heisler et al.\ (1997).  van Zee (2000) has observed 
DDO~6 ($=$ UGCA 15) and did not detect any HII regions, but did detect
a small amount of diffuse H$\alpha$ emission.  van Zee (2000) also observed
DDO~226 ($=$ UGCA 9) and did detect the presence of faint HII regions.
Jerjen et al.\ (1998) detected a faint HII region in ESO~294-G010.
ESO~410-G05 is an HI non-detection, and, based on HST imaging, 
Karachentsev et al.\ (2000) have determined that this is a dSph galaxy.  
In light of the results of the more recent deeper spectroscopy, 
it might be interesting to re-observe the other 
non-detection by Miller (1996), i.e., UGCA 438 ($=$ ESO~407-G18).
Nonetheless, the claim by Miller (1996) 
that the average current star formation for dIs in the Sculptor group is
suppressed relative to other nearby groups appears to be confirmed
(see discussion in \S 3).

\section {The HII Regions and the Star Formation Rates the Sculptor Group
Dwarf Irregular Galaxies}

\subsection{The HII Region Luminosities}

As typical for dI galaxies, the majority of HII regions in these dwarfs 
have a random asymmetric distribution throughout the galaxies (Brosch et al.\ 
1998; van Zee 2000).  However, the most luminous HII regions are found at 
or near the center (in NGC~625, NGC~59 and ESO~347-G17).  H$\alpha$ 
luminosities were calculated from the H$\alpha$ fluxes using the distances 
of Table 1 and assuming a Galactic extinction correction of the form: 
A(H$\alpha $) = 2.32 $E(B-V)$ (Miller \& Hodge 1994), using reddening values 
from Schlegel et al.\ (1998).  Figure 2 shows a histogram of the HII region 
luminosities for all the dwarfs.  The bulk of them are mostly low luminosity, 
around 10$^{37.5}$ erg s$^{-1}$, very similar to what is found in other 
nearby irregulars (Youngblood \& Hunter 1999) and isolated dIs 
(van Zee 2000).  The lowest luminosity HII region, of 10$^{35.4}$ erg s$^{-1}$,
is found in SC~24, the nearest dwarf in our sample and also the lowest 
luminosity dwarf of the group.  For comparison, the Orion nebula has
an H$\alpha$ luminosity of 10$^{37}$ erg s$^{-1}$ and 30 Doradus in the LMC
has an H$\alpha$ luminosity of 1.5 $\times$ 10$^{40}$ erg s$^{-1}$ 
(Kennicutt 1984).

The most luminous HII regions  are found in NGC~625, which 
is obviously undergoing an active phase of star formation, with its central 
region covered with bright HII regions. Its most extreme region (\#5 in 
Figure 1d) is actually a blend of at least 2 peaks (barely resolved with our
image quality of 1\arcsec), and the H$\alpha$ luminosity of the whole complex 
(4.5 $\times$ 10$^{39}$) is approaching that of the supergiant HII region
30 Doradus.  HII regions with these high luminosities are characteristic of
nearby BCDs and are even found in a few irregulars as well (e.g., Youngblood 
\& Hunter 1999). In NGC~625, several arcs and filaments, typical signs of 
active star formation, are seen around the larger brighter complexes 
(regions \#4 and \#8 for example).  Most interesting are the two spectacular 
`chimney'-like radial filaments, extending out of the plane of the galaxy 
perpendicular to the major axis (regions \#11 and \#12).  These structures 
probably trace the conduits of outflows of hot gas upward into the 
halo from the inner star-forming regions.  The two `chimney'-filaments 
extend to about 300 pc in length, in a galaxy with a total diameter 
(at $\mu _B$=25 mag arcsec$^{-2}$) of 3.6 kpc. 
This is larger than similar chimneys 
traced in other galaxies with sites of intense star-formation, e.g., 
the irregular NGC~4449 (Bomans  et al.\ 1997) or those seen in the 
Galaxy (Normandeau et al.\ 1996).  It is expected that in low mass 
galaxies such as NGC~625, the conditions for massive outflows of hot gas 
from galactic winds should be optimized, and, indeed, Bomans \& Grant (1998)
detected x-ray emission from NGC~625 with the ROSAT satellite observatory.
One would expect a non-negligible 
fraction of the ISM to be driven out of the galaxy.  High-resolution HI maps 
of NGC~625 show very unusual HI kinematics, with the gas in large-scale 
rotation around the major axis rather than the minor one (C\^ot\'e et al.\ 
2000), although no corresponding features at the location of the chimneys 
are seen (with the beam resolution of 15\arcsec).  Marlowe et al.\ (1997)
did {\it not} detect line-splitting in H$\alpha$ (a signature of expanding
bubbles), but their slit position did not cover the filaments labeled
regions \#11 and \#12 (and it may be that the gas is preferentially
expanding along the morphological minor axis, which is perpendicular
to the line of sight).
 
\subsection{Global Star Formation Rates and Timescales}

Following Kennicutt, Tamblyn, \& Congdon (1994), H$\alpha$ luminosities 
were converted to current SFRs with:
\begin{equation}
SFR(total) = {{L(H\alpha )}\over{1.26\times 10^{41} erg\  s^{-1}}}\ 
 M_\odot\  yr^{-1}
\end{equation}
which has been derived 
for normal spiral galaxies with a modified Salpeter IMF.  This is slightly 
different ($\sim$ 10\% lower) from the conversion proposed by Kennicutt (1983)
and is the conversion adopted by Kennicutt (1998) in his grand synthesis of
global SFRs in galaxies.  We calculated the 
rates in our dwarf galaxies in the same way although no systematic adjustment
was made for internal extinction (as for spiral galaxies in Kennicutt 1998)
since there is a wealth of evidence that extinction is usually quite small
in these low metallicity systems. 

Note that using a conversion from H$\alpha$ luminosity to SFR derived for 
normal spiral galaxies and applying it to dwarf galaxies is liable to 
both increased uncertainties and possible biases.  The main sources of
uncertainty arise from metallicity dependences of both the IMF and the
production of ionizing photons by the stars.  However, our main purpose here
is to compare the present sample of dwarf galaxies to other samples of 
dwarf galaxies, and thus, it is
most important to be consistent in the derivation of SFRs so that we
can compare directly to other studies like that of van Zee (2001).
The possible larger uncertainties and biases for the present sample
will be similar to those of other studies of dwarf galaxies.
Additionally, H$\alpha$ measurements of star formation rates are
measuring massive star formation rates and carry an implicit assumption
of a constant IMF. 

Table 3 shows the results, where 
the H$\alpha$ luminosities used in Equation (2) are the sum of all the HII 
region luminosities in each galaxy.  Summing over just the HII regions may
underestimate the total H$\alpha$ luminosities by neglecting diffuse 
H$\alpha$ emission that ``leaks out'' of the HII regions (Hunter, Hawley,
\& Gallagher 1993; Ferguson et al.\ 1996).  Youngblood \& Hunter (1999) 
and van Zee (2000) note that,
on average, the H$\alpha$ fluxes from the sums of individual HII regions
underestimate the total H$\alpha$ emission by roughly a factor of two.
However, van Zee (2000) points out that in low surface brightness (LSB) galaxies
with low HII region coverage fractions, large aperture H$\alpha$ measurements
are susceptible to significant errors with even small continuum 
subtraction errors.  Thus, in the present paper, we have chosen to 
simply report the sum of the H$\alpha$ emission from the HII regions.

These SFRs vary over 
several orders of magnitude but most of them are extremely low, as the dwarfs 
of Sculptor are only experiencing modest star-forming activity.  The SFR 
is known to be a strong function of absolute magnitude, for example, in the 
dwarf galaxies of the M81 Group (Miller \& Hodge 1994), and thus we have
calculated SFRs normalized to blue luminosities in Table 3.  Normalizing
to L(B) significantly decreases the range, and the extreme cases on
the low end (ESO~348-G09 and SC~24) and the high side (ESO~473-G24 and 
NGC~625) are now easily recognized.  The rates for the Sculptor dIs are 
consistent with those of the M81 dIs for a given magnitude, and our 
values here for extremely faint dwarfs like SC~18 and SC~24 
extend this trend to fainter levels.  

An interesting way to characterize the present SFR is 
to compare it to the total gas mass and estimate a gas depletion time scale,
($\tau_{gas}$), the number of years that a galaxy may continue to form
stars at the current rate.  HI gas masses and values of $\tau_{gas}$
are listed in Table 3.   As pointed out by Kennicutt et al.\ (1994),
$\tau_{gas}$ is really a lower limit because it does not account for
the gas returned to the ISM by natural stellar evolution processes.
Nonetheless, the quantity $\tau_{gas}$ is an interesting one, and 
the range in $\tau_{gas}$ for the Sculptor Group dIs is quite impressive.  
The currently star bursting NGC~625 shows a gas consumption time scale of
only 3 Gyr, while several of the galaxies have consumption time scales
on order or in excess of 10 times the present age of the Universe.

Our first calculation of the gas consumption time scale for  
the extremely quiescent galaxy SC~24 resulted in a value of
4800 Gyr, but we suspected that this was most likely due to an 
overestimate of its total HI gas mass.  The width of the HI emission 
line is listed by C\^ot\'e et al.\ (1997) as $\Delta$V$_{20}$ $=$ 
92 km~s$^{-1}$, which would be an unusually large line width for such 
a low luminosity galaxy.  The value of M(HI)/L(B) (23.2) was also 
unusually large.  Inspection of the 21 cm spectrum published by
C\^ot\'e et al.\ (1997) showed that the detection is close to 
the range of Galactic emission (with a heliocentric velocity of 
only 79 km s$^{-1}$) and much of the HI flux could be due to curvature
in the continuum baseline that is not well fit.  We inspected the 
HIPASS data base (Barnes et al.\ 2001) and found an HI detection
at the position of SC~24 at a velocity of $\approx$ 80 km~s$^{-1}$, 
but with a $\Delta$V$_{20}$ $\approx$ 25 km~s$^{-1}$, and thus a
flux integral of only 3.2 Jy km~s$^{-1}$ (roughly one 
quarter that of the original C\^ot\'e et al.\ (1997) value).  Here we 
will use the HIPASS value, but interferometric HI
observations of SC~24 will be necessary to truly resolve this issue.  
Note that with the new, lower value for the flux integral, 
M(HI)/L(B) ($=$ 6.29) is still high, and the gas consumption time scale 
of 1300 Gyr is still very large.
 
Figure 3 shows histogram comparisons of the values of $\tau_{gas}$ 
for the Sculptor Group dIs with three comparison groups,
the Local Group dIs (from Mateo 1998), the gas-rich LSB
galaxies studied by van Zee et al.\ (1997), and the
isolated dIs of van Zee (2000, 2001).
In comparison to the Local Group dIs, the distribution of 
Sculptor Group dIs shows a shift to larger values of $\tau_{gas}$ (although
there is considerable overlap between the two samples).
The range for both the Sculptor Group and Local Group samples
is large when compared to that seen for the gas-rich quiescent 
LSB dIs and the isolated dIs.  This is due, at 
least in part, to selection effects.   The two nearby group samples
have a higher proportion of extremely low luminosity galaxies,
and the smaller galaxies are much more liable to large relative
variations in current SFR, thus resulting in a large range in 
$\tau_{gas}$.  Large gas content played a primary role in the
selection of the LSB dIs, and a secondary role in the selection of
the isolated dIs, so a bias toward larger values of $\tau_{gas}$
in these samples is expected.

Are the Sculptor Group dIs really
that different from the Local Group dIs? 
The Local Group has two galaxies with values of log ($\tau_{gas}$) 
$\le$ 9, and the Sculptor Group has three galaxies
with log ($\tau_{gas}$) $\ge$ 11.5, so, at least statistically
these two distributions could be found to be different.
Using the Kolmogorov-Smirnov (hereafter K-S) test (specifically 
{\it kstwo} of Press
et al.\ 1992) to determine the significance of the differences
between the distributions in Figure 3, we find that the
probability that the Sculptor Group and Local Group galaxies
are drawn from the same distributions is 30\%.  This emphasizes
the large degree of overlap in the two samples.  Interestingly,
the K-S test yields a probability of 89\% that the Sculptor Group
and LSB samples are drawn from the same distribution.  However, this
reflects the fact that the K-S test has a higher sensitivity to the
median values of distributions than to the spreads in distributions
(Press et al.\ 1992); the standard deviations of the two samples 
differ by a factor of two. The F-test statistic ({\it ftest} from Press
et al.\ 1992) indicates that the Sculptor Group and the LSB dIs
have significantly different variances (significance of 99\%).
All other K-S test comparisons of the samples shown in Figure 3 
yield probabilities of less than 10\%.  
Note that since the current star formation rates represent a snapshot
in time, we have no guarantee that these distributions might
not have looked very different just a few 10$^8$ years ago.

Another important caveat should be added at this point; missing 
from Figure 3 are gas-rich galaxies with current star formation
rates of zero (no H$\alpha$ emission).  The Local Group has
five of these dIrr/dSph or ``transition'' galaxies (Mateo 1998)
and the Sculptor Group has three.  Although not plotted in Figure
3, the consideration of these galaxies would increase the 
degree of overlap in the two samples.  The transition galaxies
are discussed in more detail in \S 4.    
 
There are different physical reasons for a galaxy to have a very large
value of $\tau_{gas}$.  For example, van Zee et al.\ (1997) and
Kennicutt \& Skillman (2001) found values in excess of 200 Gyr for the
LSB dI DDO~154.  This appears to be mainly due to
the very large HI halo surrounding this galaxy which contains 90\% of
the galaxy's HI.  In this case, most of this gas is probably unavailable
for star formation.  In the case of episodic star formation, it is
possible to observe a galaxy with a current SFR that
is well below its average SFR.  This results in an 
artificially large value of $\tau_{gas}$.  With detailed recent star
formation histories like those available from imaging studies of the
young stellar population (e.g., Dohm-Palmer et al.\ 1997, 1998) it is
possible to get a better measure of the {\it average} recent SFR.

Often the H$\alpha$ and total luminosities of galaxies are used in
order to investigate how the current SFR compares to some
measure of the
past average SFR.  This ratio has been referred to as
both the ``star formation timescale'' (Roberts 1963, Hodge 1993) which
is the ratio of the mass of the stars present to the current rate of
star formation and the ``birthrate parameter'' (Scalo 1986,
Kennicutt et al.\ 1994) which is the ratio of the past average SFR
to the current SFR.
Deriving the past average SFR requires estimating
the total mass of the stars formed over the lifetime of the galaxy.
This can be done by converting the color of the stellar
population into a representative mass-to-light ratio (e.g.,
Miller 1996).  Alternatively, Kennicutt et al.\ (1994) convert the
H$\alpha$ equivalent width directly into a birthrate parameter.
Both methods depend on assumptions of stellar and galaxy evolution,
and making assumptions about galaxy evolution in the study of galaxy 
evolution may produce misleading results
(see discussion in van Zee 2001).  Often these models are constructed
with large, solar metallicity spiral galaxies in mind, so the direct
applicability to low metallicity dwarfs carries an additional
uncertainty.

Here we will follow Hodge (1993) and
make the simple assumption of a mass-to-blue light ratio of 1 so that the
ratio of the current SFR to the past average SFR
($\tau_{form}$) is directly comparable to the values
listed in Hodge (1993).
Assuming a single value of M/L for all of the galaxies is partially
supported by the small range in colors observed for these galaxies.
From the photometry of C\^ot\'e (1995), we find that all but two of
the galaxies listed in Table 3 have values of B$-$R between 0.7 and 1.0
(with ESO~471-G06 slightly bluer at 0.6 and ESO~245-G05 even bluer
at 0.3).
Thus, $\tau_{form}$ is simply L(B)/SFR with
both given in solar units.  This is listed in Table 3, column 4, which
is simply the inverse of the values in column 3.  Hodge (1993)
studied the sample of galaxies in Kennicutt (1983) and found average values
of $\tau_{form}$ of about 60 Gyr for early type spirals, 15 Gyr for
later type spirals and 8 Gyr for irregular galaxies.  The large values of
$\tau_{form}$ found here indicate that the typical dI in the Sculptor
Group is currently forming stars at a much lower rate than it has in the
past, in support of the earlier finding by Miller (1996).

Figure 4 shows histogram comparisons of the values of $\tau_{form}$
for the Sculptor Group dIs with the same three comparison groups
used in Figure 3.
As in Figure 3 where $\tau_{gas}$ is compared, the range for 
$\tau_{form}$ in the Sculptor Group dI sample is shifted to 
larger values relative to the Local Group dI sample (again
with considerable overlap between the two samples).
The Local Group
sample stands out with three galaxies with relatively low values of
$\tau_{form}$ (IC~10, GR~8, and NGC~6822).  Once again, this is due, in part, to 
selection effects.   The gas-rich quiescent LSB 
dI sample and the isolated dI sample excluded galaxies with 
exceptionally high values of current SFR (HII galaxies or 
blue compact dwarf galaxies).  It is also interesting to note
the suggestion that IC~10 should be considered
to be a blue compact galaxy (Richer et al.\ 2001 and references therein). 

Are the Sculptor Group and Local Group dI distributions different in
Figure 4?
Here the K-S test gives a probability of only 10\% that the Local 
Group and Sculptor Group samples are drawn from the same distribution.
This is due, in part, to the presence of IC~10 and GR~8 in the
Local Group sample.  If these two galaxies are deleted from the Local Group
sample, then the K-S test probability rises to 22\%, 
on the margin of a significant difference.  However, IC~10 and GR~8 do exist,
so it makes no sense to delete them from the sample.
The K-S test yields a 71\% probability for the Sculptor Group sample
and the LSB sample to be drawn from the same distribution.
The Sculptor Group and LSB samples are very similar with regard to 
mean values of both $\tau_{gas}$ and $\tau_{form}$, while the
the Local Group and isolated dI samples are also similar to each other
in these two quantities.

Interestingly, the K-S test of the Figure 4 data reveals that the 
Local Group dI sample has a 51\% probability of being drawn from the 
same distribution as the isolated dI sample, and a 15\% probability 
when compared to the LSB dI sample.  These are significantly higher than 
the respective $\tau_{gas}$ comparisons (8\% and 5\% respectively).
In fact, the Local Group and isolated dI sample distributions in
Figure 4 do not look that similar, and the high significance once again 
reflects the fact that the K-S test has a higher sensitivity to the
median values of distributions than to the spreads in distributions.

In Figure 5 we compare the two timescale values ($\tau_{gas}$
and $\tau_{form}$) to each other for the four samples.
Figure 5 shows a general correlation of  $\tau_{gas}$ with 
$\tau_{form}$ (as should be expected since both axes share a 
common denominator).  The main reason for the relatively good
correlation in Figure 5 is that the gas-rich dwarf galaxies generally
show a limited range in M(HI)/L(B) (Skillman 1996).  The dotted
line in Figure 5 represents an equality between $\tau_{gas}$
and $\tau_{form}$, or, equivalently, M(HI)/L(B) $=$ 1.0. 
The conclusions of Figure 3 and 4 are
supported.  While the Sculptor Group dIs and the Local
Group dIs tend to fall into the region of the graph that is
well populated by the larger LSB and isolated dI samples, it
is interesting that the Sculptor group dIs lie at preferentially
higher values of log($\tau_{gas}$) with the majority above 10.5, while
the Local Group dIs lie at preferentially lower values of log($\tau_{gas}$)
with the majority below 10.5.  It is interesting that, when normalized
by SFR, the LSB galaxies do not deviate far from the trend.  Naively,
one might expect them to deviate to the upper right (higher gas
content and lower luminosity), but, in fact, they do not appear
to distinguish themselves in this diagnostic diagram.  This is simply
due to the fact that the dynamic range in the SFR is much larger than
the dynamic range in M(HI)/L(B).

Finally, in Figure 6, we have plotted the SFR versus gas mass 
with both normalized to the galaxy luminosity (SFR/L$_B$ versus 
M$_{HI}$/L$_B$)\footnotemark .  In this 
diagnostic diagram excursions are much more prominent.  The high
relative SFR galaxies from the Local Group are not unusually gas
rich.  The Sculptor Group galaxies tend toward larger gas fractions
and lower values of SFR (note the large number of them in the 
lower right quadrant).  The isolated dIs show a large range in values,
reaching toward the extremes seen in the two group samples.  It would
appear that the isolated dIs sample of van Zee (2000, 2001) covers the
range of properties of dIs with the possible exception of the lowest
luminosity systems.
\footnotetext{The value of M(HI)/L(B) for SagDIG in Table 4 of Mateo 
(1998) of 8.6 is in error, and we use the corrected value of 0.86 here.}

\section{Comparing ``Transition'' Galaxies and dI Galaxies}

Of course, the very large values of $\tau_{gas}$ seen in several of the 
Sculptor Dwarf dIs do not represent an upper bound; there are several Sculptor
Group dIs with HI detected but no H$\alpha$ emission detected, and
these would have $\tau_{gas}$ values of infinity.  Such galaxies are
not limited to the Sculptor Group; in the Local Group we have LGS-3,
Antlia, and DDO~210 as examples of dIs with detectable HI but no
detectable H$\alpha$ emission (Mateo 1998).  These galaxies are often
referred to as ``transition'' galaxies or dIrr/dSph galaxies\footnotemark .
\footnotetext{Note that this definition of ``transition'' dwarf galaxy is
different from of Sandage \& Hoffman (1991) which was used to characterize 
galaxies whose optical appearances showed both early and late-type
dwarf characteristics.  Knezek, Sembach, \& Gallagher (1999) have 
studied the transition galaxies of Sandage \& Hoffman (1991) and
have found them to be a rather heterogeneous sample.} 
Mateo (1998) also includes Pegasus and Phoenix in this category. 
Pegasus has two detected HII regions (Skillman, Bomans, \& Kobulnicky
1997) and the formal calculation of $\tau_{gas}$ for Pegasus yields a 
value of 3220 Gyr\footnotemark.  Until recently, an optical radial velocity was 
unavailable for Phoenix, and it was not  
clear whether the HI detected in the direction of 
Phoenix was directly associated with it (Carignan, Demers, \& C\^ot\'e
1991; Oosterloo et al.\ 1996, Young \& Lo 1997, St-Germain et al.\ 1999).  
However, Gallart et al.\ (2001) have now
provided an optical radial velocity of $-$52 $\pm$ 6 km s$^{-1}$, which is
relatively close to the HI cloud separated from Phoenix by 6$\arcmin$ 
with a velocity of $-$23 km s$^{-1}$, and they conclude that the 
properties of this HI cloud are consistent with having been recently 
lost by Phoenix.  The recent measurement of the optical radial velocity of
Phoenix of $-$13 $\pm$ 9 km s$^{-1}$ by Irwin \& Tolstoy (2002) strengthens
the connection between the HI cloud and the galaxy. 
Note that the stellar population study by 
Holtzman, Smith, \& Grillmair (2000) indicates that Phoenix has
experienced star formation up until roughly 100 million years ago
which implies that Phoenix must have had some gas until very recently. 
\footnotetext{This calculation uses the H$\alpha$ flux of the HII regions.
If one uses the total H$\alpha$ flux reported by Hunter et al.\ (1993),
then the value of $\tau_{gas}$ decreases to 252 Gyr.  Values for
Pegasus are not plotted in Figures 3, 4, 5, \& 6 because the SFR 
rate listed in Mateo (1998) is 0.} 

For the galaxies with the very low values of SFR (or high values of
$\tau_{gas}$ and $\tau_{form}$), the precise value of the current SFR 
is probably not very meaningful.  The conversion from 
H$\alpha$ flux to SFR calculated by Kennicutt et al.\ (1994) is based on 
a fully populated IMF, and with so few HII regions, it is clear that the
whole range of massive stars is not represented.  Since the presence or
absence of a few HII regions can move an extremely low luminosity galaxy 
between the dI and transition categories, they may have much in
common.  One also has to consider the uncertainty introduced by the
differences between the H$\alpha$ fluxes calculated from the sum of the 
HII regions and those calculated from the entire image (including the 
diffuse component) which can be significant for these very quiescent
galaxies.  

There are at least two different evolutionary paths that a dI galaxy
can take to become a transition galaxy.  One possibility is for the galaxy
to lose enough of its cold gas to halt present star formation.
A second possibility is to have sufficient gas for star formation,
but to be simply ``in between'' episodes of star formation.
Phoenix may be a local example of the first case 
(as discussed by Gallart et al.\ 2001), while 
Antlia, DDO~210, LGS-3, and Pegasus may be examples of the second case.  
Typically, dI galaxies have M$_{HI}$/L ratios of about one
in solar units (Skillman 1996).  Thus, galaxies with similar
M$_{HI}$/L but no current star formation could be what would 
normally be called dIs, but are simply between episodes of 
galaxy formation.  Simply based on the number of stars and the
average lifetimes of HII regions, the smaller the dI galaxy, the higher 
the likelihood that it could be found in such a phase.  Note that
all five of the Local Group transition galaxies have absolute V magnitudes
near the low end of the dI luminosity function (DDO~210 $=$ $-$10.0,
Phoenix $=$ $-$10.1, LGS-3 $=$ $-$10.5, Antlia $=$ $-$10.8,
and Pegasus $=$ $-$12.9; Mateo 1998).
Of the five Local Group transition galaxies, all have significant
HI contents with values of M$_{HI}$/L$_V$ close to one (DDO~210 $=$ 2.35, 
Phoenix $=$ 0.21, LGS-3 $=$ 0.32, Antlia $=$ 0.56, and Pegasus $=$0.45; 
Mateo 1998); the value for Phoenix
represents the candidate HI cloud at $-$23 km~s$^{-1}$.
Thus, four of the transition galaxies in the Local Group are consistent
with the evolutionary scenario of temporarily interrupted star formation.

In a detailed study of LGS-3, Miller et al.\ (2001) have shown how
the SFR in the outer parts of LGS-3 has decreased faster
than the SFR in the inner parts.  If the ``active'' star forming
area of a galaxy decreases with time, then the likelihood that it 
will experience a ``transition'' phase with no current massive
star formation will increase with time.  Nonetheless, it may 
retain sufficient gas for future episodes of star formation.
 
The Sculptor Group, with three galaxies with clear transition type 
properties, provides us with additional case studies of
transition type galaxies.
SDIG, DDO~6\footnotemark, and UGCA~438 are all HII region non-detections
with HI detections and M$_{HI}$/L$_B$ values close to one (0.88, 1.4, and
2.4 respectively). ESO~294-G10 is an HI non-detection from both 
C\^ot\'e et al.\ (1997) and the HIPASS database, but has
an HII region observed (Jerjen et al.\ 1998), so it might be interesting 
to re-observe it in HI.  Additionally, four of the galaxies in Table 3 have 
values of $\tau_{gas}$ larger than 100 Gyr with values of M$_{HI}$/L$_B$ 
typical for dIs (ESO~471-G06, ESO~348-G09, SC~18, and SC~24). 
\footnotetext{Note the ambiguity for DDO~6 in that H$\alpha$ emission is
detected, but HII regions are not identified.  We have included it as
a transition galaxy because no HII regions are detected with the 
somewhat arbitrary justification that H$\alpha$ emission can arise from 
sources other than photoionization from young massive stars.} 

Although there has not been a great deal of research on transition 
type galaxies, the fact that they (and galaxies with similar properties)
turn up frequently in the Local Group (5 examples) and Sculptor Group 
(3 examples) indicates that this type of galaxy may be very common among 
the lowest luminosity, gas-rich, star forming galaxies.
It is then an interesting question whether the spatial distribution of 
transition type galaxies in our Local Group and the Sculptor Group
is different from the spatial distribution of normal dIs.
In Figure 7 we plot the spirals, dIs, and transition galaxies 
from the Local and Sculptor groups in Supergalactic coordinates
(cf.\ Jerjen et al.\ 1998, Figure~9).  The distances for the spirals
are taken from Jerjen et al.\ (1998), but we have added NGC~24
at a distance corresponding to Equation (1).  The dEs, which tend to 
cluster near the spiral galaxies, are left out for clarity.
Inspection of Figure 7 shows no obvious pattern to distinguish the
dIs from the transition galaxies.  If the low SFRs
in the transition galaxies were due to the influence of the large
spiral galaxies, we might expect to see evidence of a morphology
density relationship like that observed for the dE galaxies
(e.g., Binggeli, Tarenghi, \& Sandage 1990 and references therein), 
but none is immediately evident in Figure 7.  

However, Figure 8 shows the same galaxies in another projection,
that of the Supergalactic X,Z plane.  By coincidence, the Sculptor
Group is nearly centered on the Supergalactic Y-axis
(de Vaucouleurs 1958, 1975).  Thus, if the
Local Group -- Sculptor Group concentration is modeled as a 
cylinder like filament aligned along the Supergalactic Y-axis,
this projection shows the typical displacement from the center of
that concentration.  Figure 8 gives the appearance that the transition
galaxies show a concentration to the center of this filament 
similar to that seen by the spirals, while the dIs appear to 
have larger radial distances.

Figure 9 is a histogram of the radial distances from this axis sorted by
galaxy type.  Average radial distances and standard deviations have been
calculated for each of the types.  Here we are including the dE 
galaxies for the first time.  The distances for the Local Group dE 
galaxies are taken from Mateo (1998) and van den Bergh (2000).
The distances of the six known Sculptor Group dEs come from 
Jerjen et al.\ (1998: NGC~59, SC~22, ESO~294-G010, ESO~540-G030),
Karachentsev et al.\ (2000: ESO~410-G005), and
Jerjen \& Rejkuba (2001: ESO~540-G032).   
The histogram\footnotemark\  confirms the impression 
that the radial distributions of the transition galaxies are very 
similar to those of the spiral galaxies and dE galaxies and distinctly 
different from those of the dIs. 
\footnotetext{Note that in constructing this histogram, NGC~55 appears twice
because it is generally referred to as a late type spiral member
of the Sculptor Group (SB(s)m; de Vaucouleurs, de Vaucouleurs, \&
Corwin 1976), but listed as a dI (Irr IV; van den Bergh 1994a)
member of the Local Group by Mateo (1998).} 
If the galaxies are separated into
the Local and Sculptor groups, the offset is still present.
The main difference is the significant population of dIs more
than 1 Mpc from the cloud axis.
The bimodality of the dE distribution simply reflects the fact that 
most of these galaxies are either satellites of M~31 (with an offset
from the Supergalactic Y-axis of 0.7 Mpc) or the Milky Way (with
no offset). 

Figure 10 shows the results of the calculation of the distances of 
the dI, transition, and dE galaxies to the nearest spiral galaxy 
inspired by the results of Figures 8 and 9.
In this figure, the Sculptor Group dwarfs are noted by dots
in the histograms\footnotemark .
\footnotetext{Note that as a result of the inclusion of NGC~55 
in our list of Sculptor spirals, two of the Local Group dIs listed
by Mateo have NGC~55 as their nearest neighbor spiral, and therefore
are marked with dots.  These two galaxies are IC~5152 and UKS2323-326.
Because of its listing as both a Sculptor spiral and a Local Group
dI, NGC~55 shows up as a nearest neighbor to itself, but we have
excluded that from the histogram.}
The main difference between the dI and transition galaxies in 
Figure 10 is the lack of transition galaxies at large distances, 
with the result that
the average distance for the dIs is almost a 
factor of two larger than that of the transition galaxies.  
This is a marginally stronger result from that shown in Figure 9
(0.82/0.58 $=$ a factor of 1.4 for average radial distance from the 
Cloud axis versus 0.92/0.50 $=$ a factor of 1.9 for average 
distance to the nearest spiral).
The K-S test indicates that all three distributions in Figure 10 are
significantly different.  Comparing the transition galaxies to the 
other two samples yields only a 15\% probability that the transition 
galaxies come from the same sample as the dIs and only an 8\%
probability of coming from the same sample as the dEs.

The distances calculated for Figure 10 carry a larger degree of 
uncertainty than those in Figure 9 
since the distances to most of the Sculptor dwarfs are based on the
distance-velocity relationship in Equation (1).  Nonetheless, it is
natural to ask the question whether the transition galaxies are
found preferentially near to the spiral galaxies or to
the axis of the Local Group -- Sculptor Cloud that is defined
by the spiral galaxies.  On its own, the near identity of the means 
and standard deviations of the transition galaxy distances to the
group axis (0.58, 0.24) and the nearest spiral (0.50, 0.34)
prevent us from distinguishing which effect may be more relevant
for the evolutionary status of these galaxies.  Of course, the
results for the nearest spiral galaxy distances may change as
better distances become available for the bulk of the sample. 

In comparing the transition galaxies to the dE galaxies we find
a significant difference between the distributions in Figures 9
and 10.  The ratio of the average distance to the nearest spiral
(2.3 times larger for the transition galaxies relative to the 
dE galaxies) is much larger than the ratio of the average distance 
to the group axis (1.5 times larger for the transition galaxies).
However, this is mainly reflecting the offset of M~31 (and therefore
its companion dE galaxies) from the group axis.
Although the samples are admittedly small, the impression is that
the three dwarf morphological types are separable by mean distance
to nearest neighbor spiral galaxy.  This would support the hypothesis
that a key factor affecting dwarf morphological type is 
environmental.

There is, of course, the possibility that Figure 10 is susceptible 
to the biases of selection effects.  It is a safe bet that there 
remain low luminosity, LSB galaxies undetected
in the Sculptor Group.  Recently, most of the new low 
luminosity dEs in the Local Group have been found by searching in 
the vicinity of M~31 (e.g., Armandroff, Jacoby, \& Davies 1999),
and this might skew the distribution of Local Group dEs towards
higher proximity.
Additionally, the Sculptor dEs do
not show the same degree of proximity to the spiral galaxies as
the Local Group dEs.  
In the histogram in Figure 10, the Sculptor
Group dEs (marked by dots) account for 4 of 23 of the galaxies in 
the main clump and 2 of 4 of the outliers.  
On the other hand, the distances to the 
Sculptor Group dEs are much more uncertain, and this may introduce
some scatter.  For example, for ESO~540-G030 and ESO~540-G032, the
calculated nearest spiral neighbor (NGC~247) distances are 560 kpc and
670 kpc respectively, but the minimum projected distances are
120 kpc and 50 kpc.  Ironically, the surface brightness fluctuation
distance for ESO~540-G032 placed it very close to NGC~247, but the
newer tip of the red giant branch measurement moved it over 1 Mpc
further away, so that its nearest neighbor changed to NGC~253.
It could be that the distances are still quite uncertain, and that
all of the Sculptor dEs could lie 
in the main group in the histogram determined by the Local Group dEs.  
Hubble Space Telescope distances from tip of the red giant branch 
measurements for all of these galaxies are required to place this 
comparison on firmer ground.

\section{The Nature of ``Transition'' Galaxies}

The present data indicate that, in the local cloud, transition galaxies 
are differentiated from dIs by their, on average, closer proximity 
to spiral galaxies.  
Since transition
galaxies are preferentially found to be lower luminosity systems
than the average dI galaxy, a luminosity -- distance relationship
for the dIs might cause such difference.  
Indeed, Armandroff et al.\ (1999) show that the extreme low luminosity
dSph galaxies in the Local Group are found preferentially close to our
Milky Way galaxy.
However, this does not appear to be the case for the Local Group dIs, 
since the Local Group has several relatively
low luminosity dIs at large distances ($\ge$ 1 Mpc) from the 
nearest spiral galaxies.  This points to an underlying cause
more closely related to environment (something is happening to 
these galaxies) as opposed to initial conditions (it is not
true that smaller dI galaxies are formed closer to spirals). 

What can be made of the
morphology-density relationship for the transition galaxies?
The transition galaxies 
do {\it not} show the same degree of morphology-density relationship 
as the dwarf spheroidals (i.e., the typical transition galaxy is 
not a close ($\le$ 200 kpc) companion to a large spiral galaxy).
Therefore, explanations of the morphology-density relationship
applicable to the dwarf spheroidals are not necessarily appropriate 
for the transition galaxies. 
Because there is a large degree of overlap in the spatial distributions
of the transition galaxies and the dIs, it is likely that the typical
transition galaxy represents the extreme in the distribution of low 
SFRs of quiescent dI galaxies.  The
observation that the transition galaxies tend toward the center
of the filamentary group structure, and lie closer, on average,
to the spiral galaxies, may be telling us that environmental
effects have driven otherwise normal dI galaxies to the observed
low values of SFR.

Perhaps these transition galaxies have a higher probability for
``tidal stirring'' or ``harassment'' than the typical dI, resulting
in their partial evolution to the dwarf spheroidal state
(Mayer et al.\ 2001a,b).  Although such interactions result in
a temporarily elevated SFR, the main long-term effect of these
interactions is to strip the galaxy of gas, resulting in lower
SFRs.  Given the small samples involved, and
the limited range in environment that has been sampled here,
definitive statements cannot be made.  It is not clear whether
galaxy luminosity (small galaxies are most easily found in a
quiescent state), position relative to the ``cloud'' structure (galaxies
near the centers of large scale structures are most likely to 
have the properties of transition galaxies), or position relative 
to the nearest spiral galaxy (galaxies nearest to spiral galaxies are 
most likely to be tidally stirred) is the dominant
variable in determining that a gas-rich dwarf has the properties
of a transition galaxy.   Similar studies of other environments 
should provide additional constraints on this problem.
Additionally, clearer observational predictions from the models
would be helpful.  Can tidally stirred galaxies be differentiated
from isolated galaxies by bursty star formation histories or 
strong stellar population radial gradients?  Can a burst of
star formation triggered by an external influence be distinguished
from a burst with an internal trigger?

Finally, while our conclusion that the transition galaxies
have more in common with the dI galaxies than with the dE galaxies
may be debatable, it is clear that transition galaxies are
not dE galaxies.  The similarities between transition galaxies
and dI galaxies in gas content and spatial distribution may make it 
natural to group the transition galaxies with the dIs in making 
comparative family studies.  
Note that this
is the opposite of the assumption of Blitz \& Robishaw (2000)
in their study of ``gas-rich dwarf spheroidals.''  By including
the transition galaxies in with the dwarf spheroidals, they 
concluded that a large fraction of the dwarf spheroidals in the
Local Group are associated with HI gas.  We would argue that
the transition galaxies should be excluded from this sample,
leading to the opposite conclusion, that only a very small fraction
(if any)
of dwarf spheroidals in the Local Group are associated with HI gas.  
 
\section{Conclusions}

  We have presented H$\alpha$ imaging of 
dI galaxies in the nearby Sculptor Group.  
Comparing the Sculptor Group dIs to the Local Group dIs, in support
of Miller (1996), we
find that the Sculptor Group dIs have, on average, lower values of
SFR when normalized to either galaxy luminosity or gas mass (although
there is considerable overlap between the two samples).
The range for both the Sculptor Group and Local Group samples
is large when compared to that seen for the sample of gas-rich quiescent
LSB dIs from van Zee et al.\ (1997) and the sample of
isolated dIs from van Zee (2000, 2001).  This is probably best understood
as a selection effect since the nearby group samples have a much larger
fraction of extremely low luminosity galaxies
and the smaller galaxies are much more liable to large relative
variations in current SFR.  
The Sculptor Group and LSB samples are very similar with regard to
mean values of both $\tau_{gas}$ and $\tau_{form}$, 
and the the Local Group and isolated dI samples are also similar 
to each other in these two quantities.

The properties of ``transition'' galaxies in Sculptor and the Local
Group are also compared and found to be similar.   The transition galaxies
are preferentially among the lowest luminosities of the gas rich dwarf
galaxies.  Relative to the dwarf irregular galaxies, the transition 
galaxies are found preferentially nearer to spiral galaxies, and are
found nearer to the center of the mass distribution in the local cloud.
However, the transition galaxies are not found as close to the nearest
neighbor spiral galaxies as the dE galaxies.
Most of these systems are consistent with normal dI galaxies which 
currently exhibit temporarily interrupted star formation.  The observed
density-morphology relationship indicates that environmental
processes such as ``tidal stirring'' may play a role in causing their
lower SFRs.

\acknowledgments

Special thanks are extended to Chris Lidman for acquiring the ESO~473-G24 
H$\alpha$ observations, and to Liese van Zee for the use of her astrometry 
package to derive the HII region positions.
We wish to thank G.\ Bothun,  R.\ Kennicutt, 
E.\ Tolstoy, and L. van Zee for many helpful conversations.  
John Cannon, Henry Lee, and Liese van Zee proofread and provided 
valuable comments on this manuscript. 
We also thank the referee for a prompt and careful reading of the
manuscript and several valuable comments. 
This research has made use of the NASA/IPAC Extragalactic Database 
(NED) which is operated by the Jet Propulsion Laboratory, California
Institute of Technology, under contract with the National Aeronautics 
and Space Administration.
This research has made use of NASA's Astrophysics Data System
Abstract Service. 
EDS acknowledges partial support from a NASA LTSARP grant No. NAG5-9221 
and the University of Minnesota. 
BWM is supported by the Gemini Observatory, which is operated by the 
Association of Universities for Research in Astronomy, Inc., on behalf 
of the international Gemini partnership of Argentina, Australia, Brazil, 
Canada, Chile, the United Kingdom, and the United States of America. 
We would like to dedicate this paper to the memory of Robert A.\ Schommer, 
whose many contributions to the astronomical community, and the Chilean
astronomical community in particular, will always be appreciated. 

\clearpage


\clearpage

\begin{deluxetable}{lcccccc}
\tablenum{1}
\tablewidth{0pt}
\tablecaption{Sculptor Group Dwarf Irregular Galaxies with H$\alpha$ Emission
\label{tbl-1}}
\tablehead{
\colhead{Galaxy Name} & \colhead{Alternate Name} & \colhead{R.A. (J2000)} & 
\colhead{dec.\ (J2000)} & \colhead{v$_{\rm GSR}$} & \colhead{D (Mpc)} & \colhead{M(B)}
}
\startdata
ESO~347-G17\tablenotemark{a}
            &         &23:26:57 &$-$37:20:47&695 &6.99 &$-$14.79 \\
ESO~471-G06\tablenotemark{a,b}
            &UGCA 442 &23:43:47 &$-$31:57:09&281 &3.51 &$-$14.06 \\
ESO~348-G09\tablenotemark{a}
            &         &23:49:23 &$-$37:46:19&640 &6.52 &$-$13.75 \\
SC~18       &         &00:00:59 &$-$41:09:18&119 &2.14 &$-$9.20  \\
NGC~59      &ESO~539-G04&00:15:25 &$-$21:26:36&392 &4.39 &$-$15.30 \\
ESO~473-G24\tablenotemark{a}
            &         &00:31:20 &$-$22:46:02&571 &5.94 &$-$12.68 \\
SC~24       &         &00:36:38 &$-$32:34:28& 61 &1.66 &$-$8.39  \\
DDO~226\tablenotemark{c}
            &UGCA 9   &00:43:04 &$-$22:14:49&386 &4.39 &$-$13.49 \\
            &IC 1574  &    \\
            &ESO~474-G18&  \\
DDO~6\tablenotemark{c}
            &UGCA 15  &00:49:49 &$-$21:00:54&318 &3.82 &$-$12.24 \\
            &ESO~540-G31&  \\
AM~0106-382 &         &01:08:22 &$-$38:12:33&593 &6.13 &$-$12.47 \\
NGC~625\tablenotemark{a}
            &ESO~297-G05&01:35:06 &$-$41:26:05&332 &3.93 &$-$16.31 \\
ESO~245-G05\tablenotemark{b}
            &A 143    &01:45:04 &$-$43:35:53&310 &3.75 &$-$15.59 \\

\enddata
\tablenotetext{a}{observed spectroscopically in companion paper}
\tablenotetext{b}{not imaged in H$\alpha$ here, observed 
by Miller (1996)}
\tablenotetext{c}{not imaged in H$\alpha$ here, observed 
by van Zee (2000)}
\end{deluxetable}

\clearpage

\begin{deluxetable}{lcccc}
\tablenum{2}
\tablecaption{HII Regions Positions and H$\alpha $ Fluxes \label{t:hapos}}
\tablewidth{0pt}
\tablehead{
\colhead{Galaxy \& Number} & \colhead{R.A.} & \colhead{Dec.} & 
\colhead{H$\alpha $ Flux} & \colhead{log L(H$\alpha$)} \\
\colhead{} & \multicolumn{2}{c}{(J2000)} &  
\colhead{ 10$^{-15}$ ergs cm$^{-2}$ s$^{-1}$} & \colhead{ergs s$^{-1}$}    
}
\startdata
ESO~347-G17 \#1 & 23:26:58.87 & -37:20:55.2 &  19 $\pm$ 4   & 38.1 \\ 
ESO~347-G17 \#2 & 23:26:58.34 & -37:20:56.0 & 2.1 $\pm$ 0.4 & 37.1 \\ 
ESO~347-G17 \#3 & 23:26:58.00 & -37:20:54.3 & 3.1 $\pm$ 0.6 & 37.3 \\ 
ESO~347-G17 \#4 & 23:26:57.80 & -37:20:49.4 & 1.6 $\pm$ 0.3 & 37.0 \\ 
ESO~347-G17 \#5 & 23:26:57.13 & -37:20:49.4 & 59  $\pm$ 12  & 38.6 \\ 
ESO~347-G17 \#6 & 23:26:56.41 & -37:20:52.1 & 2.3 $\pm$ 0.5 & 37.1 \\ 
ESO~347-G17 \#7 & 23:26:56.44 & -37:20:39.3 & 10.3 $\pm$ 2.1 & 37.8 \\ 
ESO~347-G17 \#8 & 23:26:55.88 & -37:20:49.6 & 4.9 $\pm$ 1.0 & 37.5 \\ 
ESO~347-G17 \#9 & 23:26:55.61 & -37:20:50.6 & 1.6 $\pm$ 0.3 & 37.0 \\ 
ESO~347-G17 \#10 & 23:26:55.15 & -37:20:51.0 & 25 $\pm$ 5   & 38.2 \\ 
ESO~347-G17 \#11 & 23:26:53.36 & -37:20:51.9 & 2.8 $\pm$ 0.6 & 37.2 \\ 
 & &  &  \\ 
ESO~348-G9 \#1 & 23:49:27.48 & -37:46:23.8 & 1.0 $\pm$ 0.2 & 36.7 \\ 
ESO~348-G9 \#2 & 23:49:27.07 & -37:46:24.9 & 0.4 $\pm$ 0.1 & 36.3 \\ 
ESO~348-G9 \#3 & 23:49:26.52 & -37:46:16.9 & 4.7 $\pm$ 0.9 & 37.4 \\ 
ESO~348-G9 \#4 & 23:49:21.67 & -37:46:21.3 & 0.7 $\pm$ 0.1 & 36.6 \\ 
 & &  &  \\ 
SC~18 \#1 & 00:00:58.02 & -41:09:24.2 & 5.3 $\pm$ 0.5 & 36.5 \\ 
 & & & \\ 
NGC~59 \#1 & 00:15:25.67 & -21:26:40.1 & 26 $\pm$ 5 & 37.8 \\ 
NGC~59 \#2 & 00:15:25.25 & -21:26:40.9 & 130 $\pm$ 26 & 38.5 \\ 
NGC~59 \#3 & 00:15:24.93 & -21:26:39.7 & 4.9 $\pm$ 1.0 & 37.1 \\ 
NGC~59 \#4 & 00:15:25.03 & -21:26:45.9 & 1.9 $\pm$ 0.4 & 36.7 \\ 
 & &  &  \\ 
ESO~473-G24 \#1 & 00:31:23.12 & -22:46:05.7 & 9.4 $\pm$ 1.9 & 37.6 \\ 
ESO~473-G24 \#2 & 00:31:22.16 & -22:46:14.6 & 12.3 $\pm$ 2.5 & 37.7 \\ 
ESO~473-G24 \#3 & 00:31:22.02 & -22:46:18.2 & 8.9 $\pm$ 1.8 & 37.6 \\ 
ESO~473-G24 \#4 & 00:31:21.28 & -22:46:14.1 & 7.0 $\pm$ 1.4 & 37.5 \\ 
 & &  &  \\ 
SC~24 \#1 & 00:36:38.35 & -32:34:45.9 & 0.8 $\pm$ 0.1 & 35.4 \\ 
 & & & \\ 
AM~0106-382 \#1 & 01:08:21.87 & -38:12:33.7 & \nodata & \nodata \\ 
AM~0106-382 \#2 & 01:08:22.57 & -38:12:35.6 & \nodata & \nodata \\ 
AM~0106-382 \#3 & 01:08:22.33 & -38:12:42.5 & \nodata & \nodata \\ 
AM~0106-382 \#4 & 01:08:21.23 & -38:12:42.6 & \nodata & \nodata \\ 
AM~0106-382 \#5 & 01:08:21.13 & -38:12:45.9 & \nodata & \nodata \\ 
 & & & \\ 
NGC~625 \#1 & 01:35:07.36 & -41:26:11.2 & 30 $\pm$ 3 & 37.8 \\ 
NGC~625 \#2 & 01:35:07.23 & -41:25:52.5 & 1.4 $\pm$ 0.1 & 36.4 \\ 
NGC~625 \#3 & 01:35:06.78 & -41:26:00.1 & 15.3 $\pm$ 1.5 & 37.5 \\ 
NGC~625 \#4 & 01:35:06.86 & -41:26:05.9 & 193 $\pm$ 19 & 38.6 \\ 
NGC~625 \#5 & 01:35:06.68 & -41:26:13.0 & 2100 $\pm$ 210 & 39.6 \\ 
NGC~625 \#6 & 01:35:06.53 & -41:26:19.8 & 51 $\pm$ 5 & 38.0 \\ 
NGC~625 \#7 & 01:35:06.34 & -41:25:52.8 & 14.1 $\pm$ 1.4 & 37.4 \\ 
NGC~625 \#8 & 01:35:06.11 & -41:26:04.2 & 23.0 $\pm$ 2.3 & 37.6 \\ 
NGC~625 \#9 & 01:35:05.85 & -41:26:11.2 & 470 $\pm$ 47 & 39.0 \\ 
NGC~625 \#10 & 01:35:05.86 & -41:25:59.6 & 22.4 $\pm$ 2.2 & 37.6 \\ 
NGC~625 \#11 & 01:35:05.73 & -41:25:44.3 & 12.0 $\pm$ 1.2 & 37.4 \\ 
NGC~625 \#12 & 01:35:05.20 & -41:25:49.4 & 7.5 $\pm$ 0.8 & 37.2 \\ 
NGC~625 \#13 & 01:35:05.36 & -41:26:07.5 & 19.3 $\pm$ 1.9 & 37.6 \\ 
NGC~625 \#14 & 01:35:05.54 & -41:26:15.5 & 32 $\pm$ 3 & 37.8 \\ 
NGC~625 \#15 & 01:35:05.26 & -41:26:12.4 & 32 $\pm$ 3 & 37.8 \\ 
NGC~625 \#16 & 01:35:05.16 & -41:26:15.1 & 14.1 $\pm$ 1.4 & 37.4 \\ 
NGC~625 \#17 & 01:35:04.59 & -41:26:18.1 & 21.1 $\pm$ 2.1 & 37.6 \\ 
NGC~625 \#18 & 01:35:04.80 & -41:26:09.3 & 161 $\pm$ 16 & 38.5 \\ 
NGC~625 \#19 & 01:35:04.64 & -41:26:01.4 & 4.7 $\pm$ 0.5 & 37.0 \\ 
NGC~625 \#20 & 01:35:04.42 & -41:26:00.9 & 2.6 $\pm$ 0.3 & 36.7 \\ 
NGC~625 \#21 & 01:35:03.99 & -41:26:07.3 & 7.6 $\pm$ 0.8 & 37.2 \\ 
NGC~625 \#22 & 01:35:01.96 & -41:26:25.8 & 23.0 $\pm$ 2.3 & 37.6 \\ 
NGC~625 \#23 & 01:35:01.47 & -41:26:25.9 & 19.3 $\pm$ 1.9 & 37.6 \\ 
\enddata
\end{deluxetable}

\clearpage

\begin{deluxetable}{lcccccc}
\tablenum{3}
\pagestyle{empty}
\tablecaption{Star Formation Properties of Sculptor Group dI Galaxies\label{tbl-4}}
\tablewidth{0pt}
\tablehead{
\colhead{Galaxy} & \colhead{SFR} & \colhead{SFR/L(B)} & 
\colhead{$\tau_{form}$} &
\colhead{M(HI)\tablenotemark{a}} & 
\colhead{M(HI)/L(B)} & \colhead{$\tau_{gas}$\tablenotemark{b}} \\
\colhead{} &   \colhead{M$_{\odot}$ yr$^{-1}$} & 
\colhead{M$_{\odot}$ yr$^{-1}$ L$_{\odot}^{-1}$} &
\colhead{Gyr} & 
\colhead{10$^6$ M$_{\odot}$}  & \colhead{M$_{\odot}$/L$_{\odot}$} & 
\colhead{Gyr} 
}
\startdata
ESO~347-G17 & $6.3\times 10^{-3}$ & $5.3\times 10^{-11}$  & 19
            & 120  & 1.00 & 25   \\ 
ESO~471-G06 & $2.0\times 10^{-3}$ & $3.3\times 10^{-11}$  & 30
            & 163  & 2.66 & 110  \\ 
ESO~348-G09 & $2.8\times 10^{-4}$ & $0.6\times 10^{-11}$  & 167
            & 84.3 & 1.83 & 400  \\ 
SC~18       & $2.4\times 10^{-5}$ & $3.4\times 10^{-11}$  & 29 
            &  5.0 & 7.13 & 270  \\ 
NGC~59      & $3.1\times 10^{-3}$ & $1.6\times 10^{-11}$  & 63
            & 16.7 & 0.87 & 7.1  \\ 
ESO~473-G24 & $1.3\times 10^{-3}$ & $7.6\times 10^{-11}$  & 13
            & 63.8 & 3.71 & 65   \\ 
SC~24       & $2.1\times 10^{-6}$ & $0.6\times 10^{-11}$  & 167
            & 2.1  & 6.29 & 1300 \\ 
DDO~226     & $1.3\times 10^{-3}$ & $3.6\times 10^{-11}$  & 28
            & 33.9 & 0.93 & 36   \\ 
DDO~6       & $2.9\times 10^{-4}$ & $2.5\times 10^{-11}$  & 40
            & 15.4 & 1.34 & 70   \\ 
NGC~625     & $5.0\times 10^{-2}$ & $10.2\times 10^{-11}$ & 9.8
            & 118 & 0.24 & 3.1  \\ 
ESO~245-G05 & $1.0\times 10^{-2}$ & $3.9\times 10^{-11}$  & 26
            & 289  & 1.15 & 38   \\ 
\enddata

\tablenotetext{a}{Total galaxy HI mass from C\^ot\'e (1995),
adjusted to the distance in Table 1} 
\tablenotetext{b}{$\tau_{gas}$ is the gas depletion time scale $=$ 
(Total Gas Mass)/(SFR),
where the total gas mass is 1.32 $\times$ M(HI) to account for He}

\end{deluxetable}

\clearpage

\begin{figure}
\vbox to7.2in{\rule{0pt}{2.6in}}

A complete gzipped .ps file containing Figure 1 can be obtained 
via anonymous ftp from ftp.astro.umn.edu /pub/users/skillman/sculptor
in paper1.ps.gz

\figcaption{Images of 8 Sculptor Group dwarf irregular galaxies.
The original H$\alpha$ images of the galaxies are shown in the left
panels and the continuum ($i$-band) subtracted H$\alpha$ images are shown in
the right panels.  The field of view for all of the images is identical
at 301 $\times$ 301 pixels or 1.96\arcmin\  $\times$ 1.96\arcmin .  
The HII regions are labeled.  Astrometric positions
and H$\alpha$ fluxes are listed in Table 2.
\label{fig1}}
\end{figure}




\clearpage 

\begin {figure}
\vbox to7.2in{\rule{0pt}{2.6in}}
\includegraphics{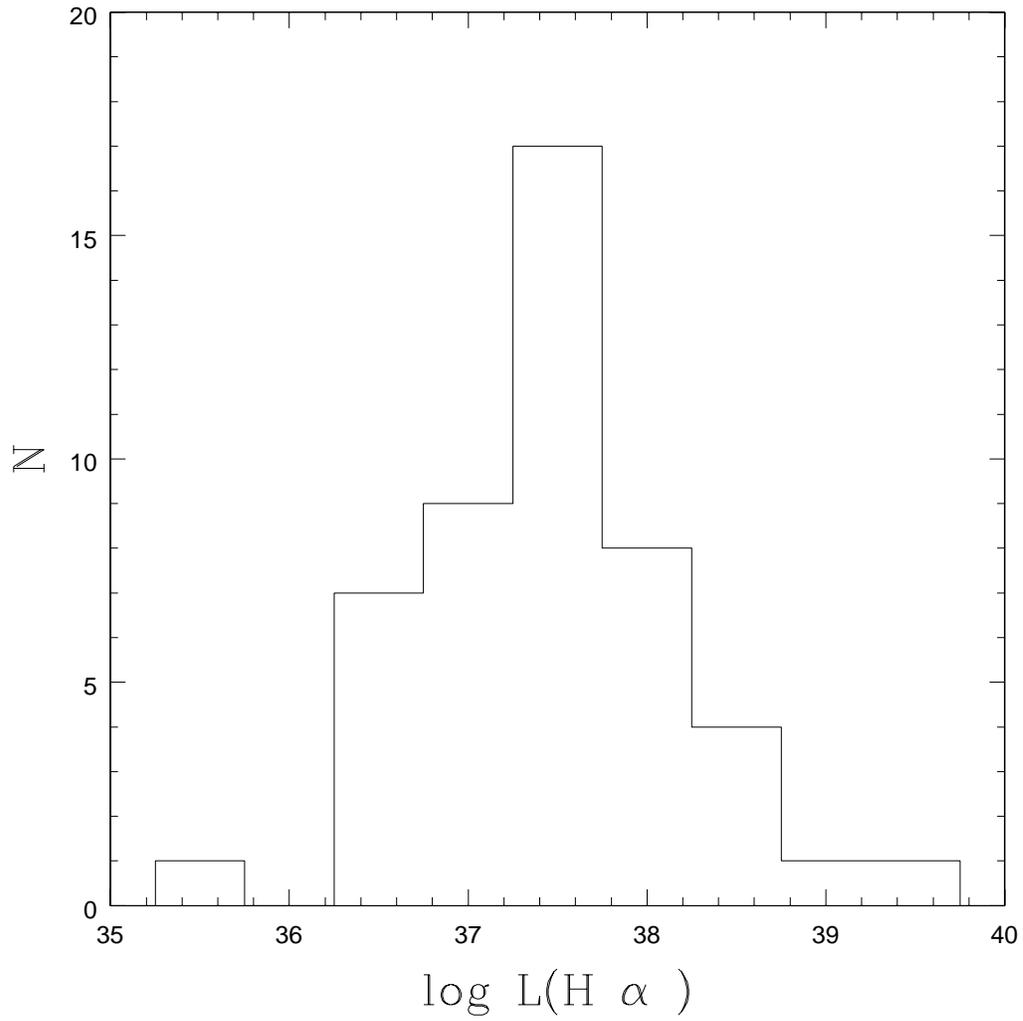}
\figcaption{A histogram of the HII region luminosities for the 8 Sculptor Group
dIs observed here in H$\alpha$.
\label{fig2}}
\end {figure}

\clearpage 

\begin {figure}
\plotone{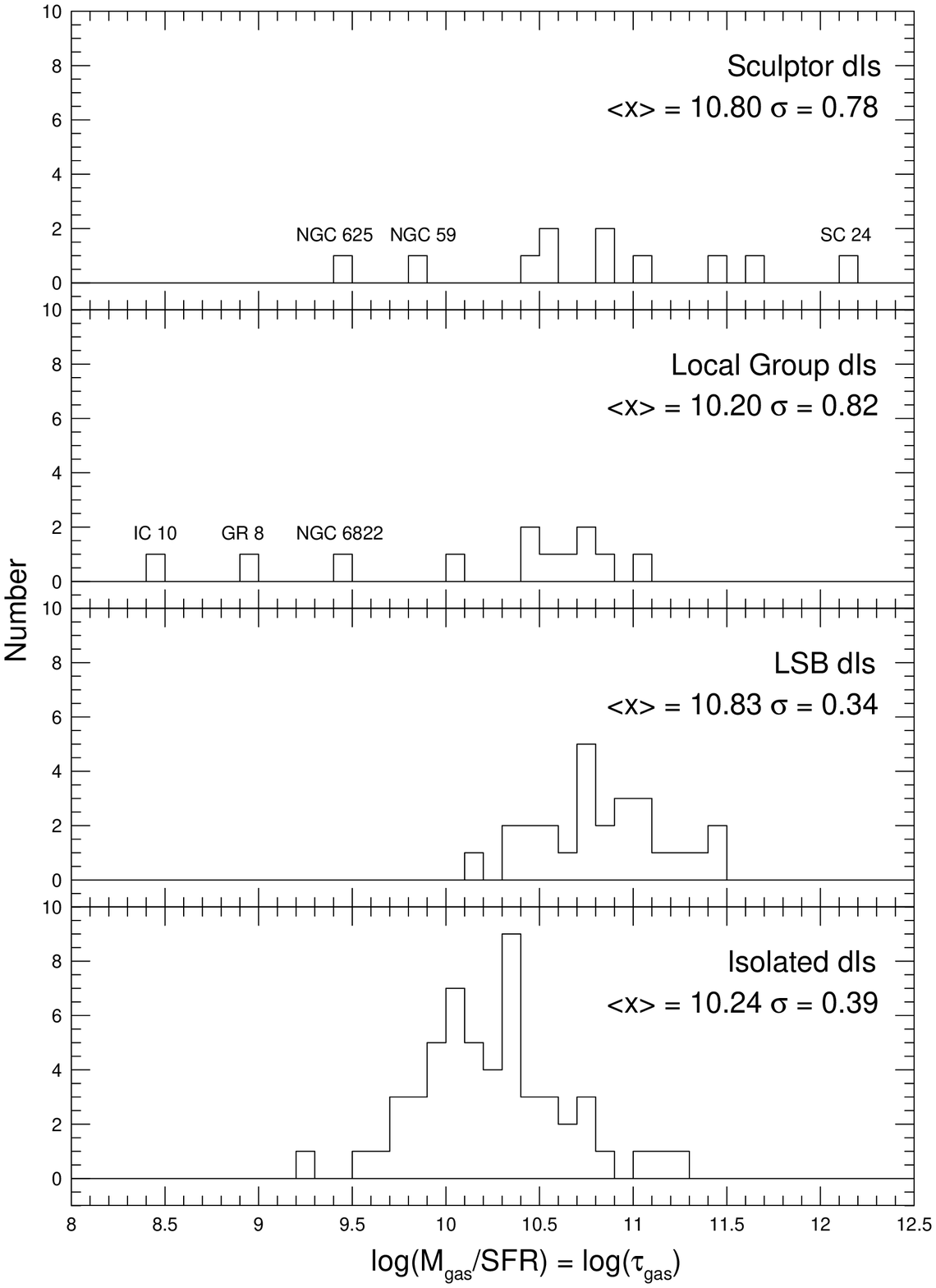}
\figcaption{
A comparison of the ratio of the gas mass to the current star formation rate
($=$ $\tau_{gas}$) for the Sculptor Group dIs and
three comparison groups:
the Local Group dIs (from Mateo 1998), the gas-rich LSB
galaxies studied by van Zee et al.\ (1997), and the
isolated dIs of van Zee (2000, 2001).
For each sample, the mean value and the standard deviation in the sample is given.
\label{fig3}}
\end{figure}

\clearpage 

\begin {figure}
\plotone{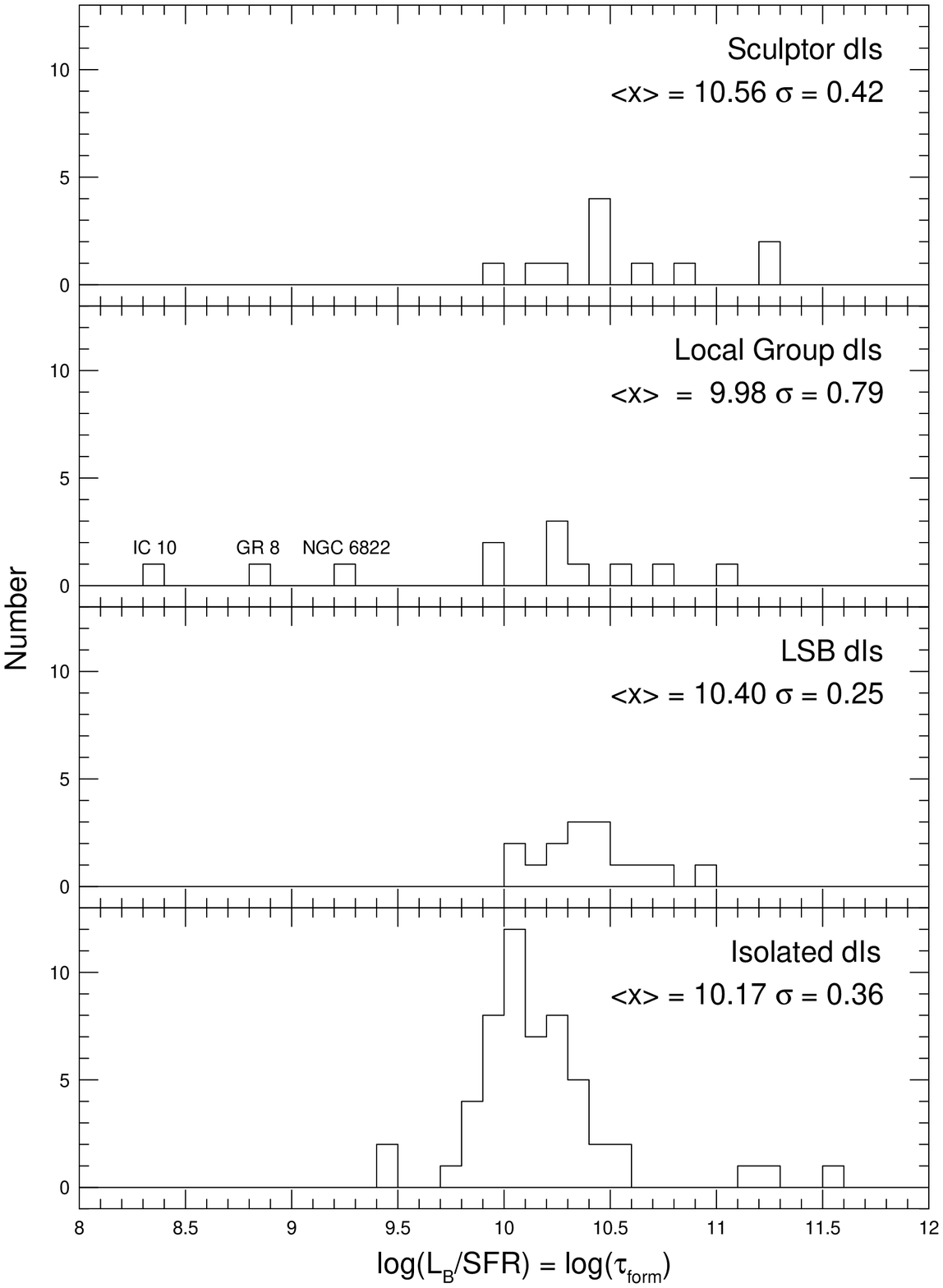}
\figcaption{
A comparison of the ratio of the luminosity to the current star formation rate
($=$ $\tau_{form}$) for the Sculptor Group dIs and
three comparison groups:
the Local Group dIs (from Mateo 1998), the gas-rich LSB
galaxies studied by van Zee et al.\ (1997), and the
isolated dIs of van Zee (2000, 2001).
For each sample, the mean value and the standard deviation in the sample is given.
\label{fig4}}
\end{figure}

\clearpage 

\begin {figure}
\plotone{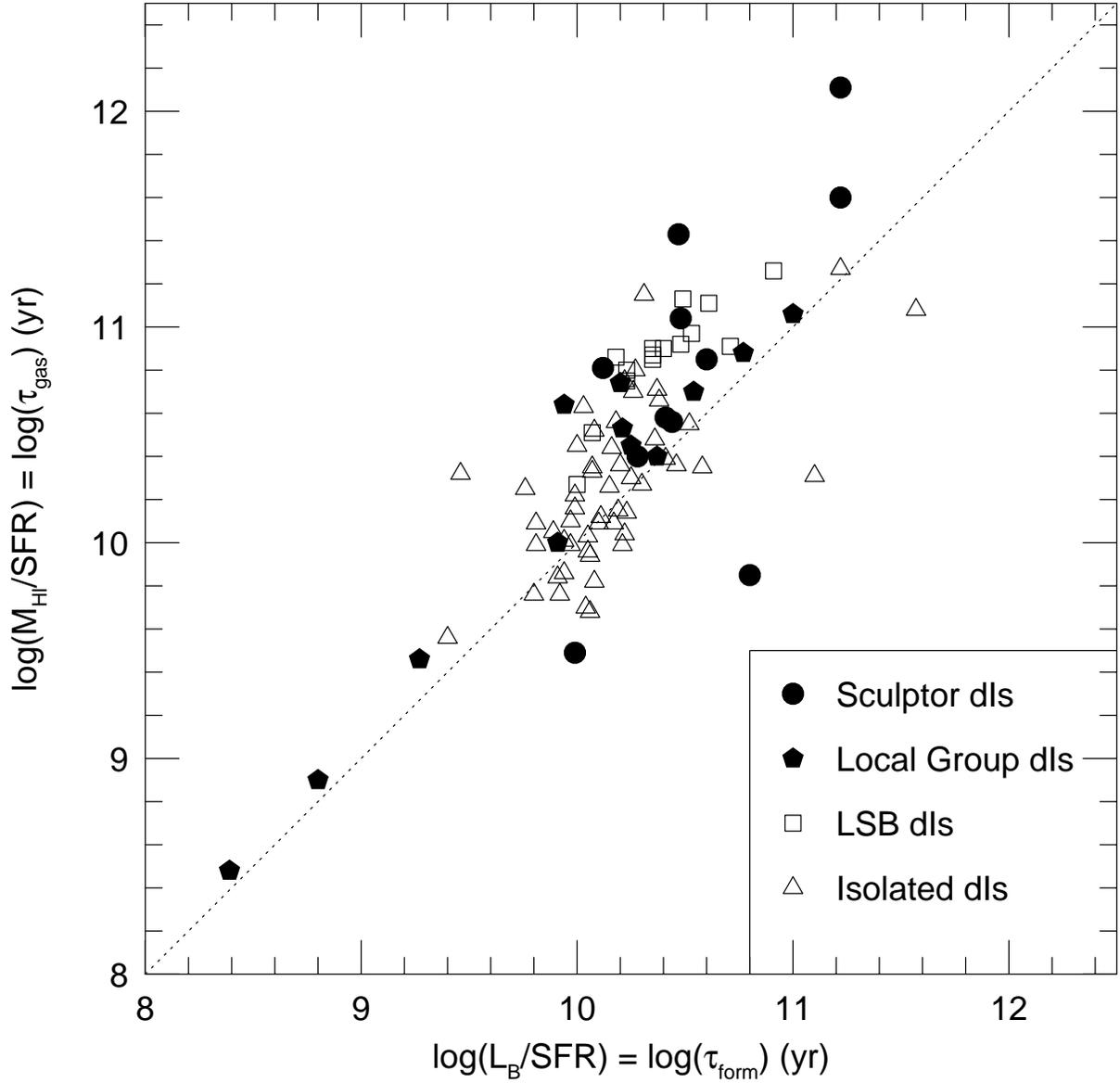}
\figcaption{
A comparison of the ratio of the gas mass to the current star formation rate
($=$ $\tau_{gas}$) to
the ratio of the luminosity to the current star formation rate
($=$ $\tau_{form}$) for the Sculptor Group dIs and
three comparison groups:
the Local Group dIs (from Mateo 1998), the gas-rich LSB
galaxies studied by van Zee et al.\ (1997), and the
isolated dIs of van Zee (2000, 2001).
The dotted line represents an equality between $\tau_{gas}$
and $\tau_{form}$, or, equivalently, M(HI)/L(B) $=$ 1.0.
\label{fig5}}
\end{figure}

\clearpage 

\begin {figure}
\plotone{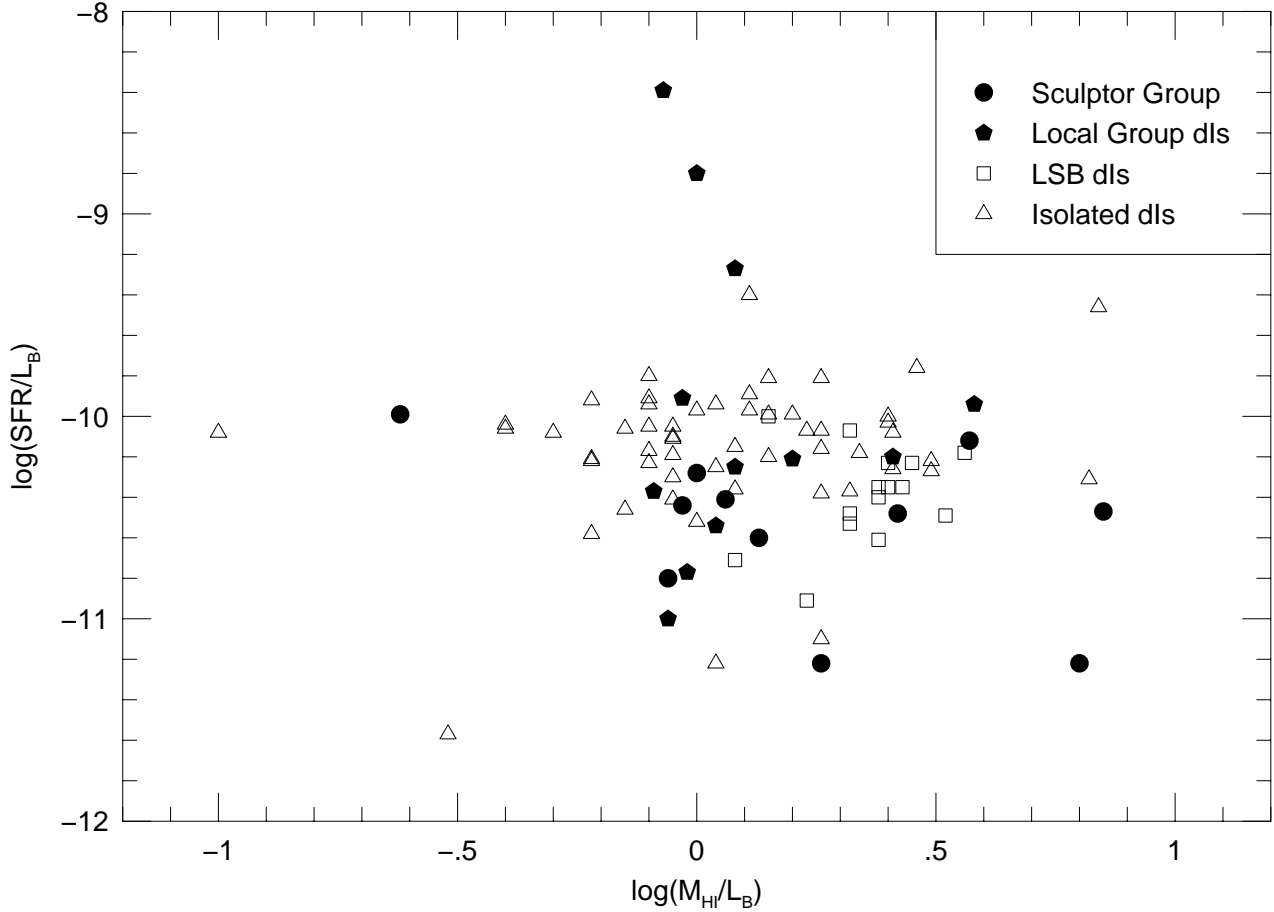}
\figcaption{
A comparison of SFR and gas mass normalized to the galaxy luminosity
the Sculptor Group dIs and
three comparison groups:
the Local Group dIs (from Mateo 1998), the gas-rich LSB
galaxies studied by van Zee et al.\ (1997), and the
isolated dIs of van Zee (2000, 2001).
\label{fig6}}
\end{figure}

\clearpage 

\begin {figure}
\plotone{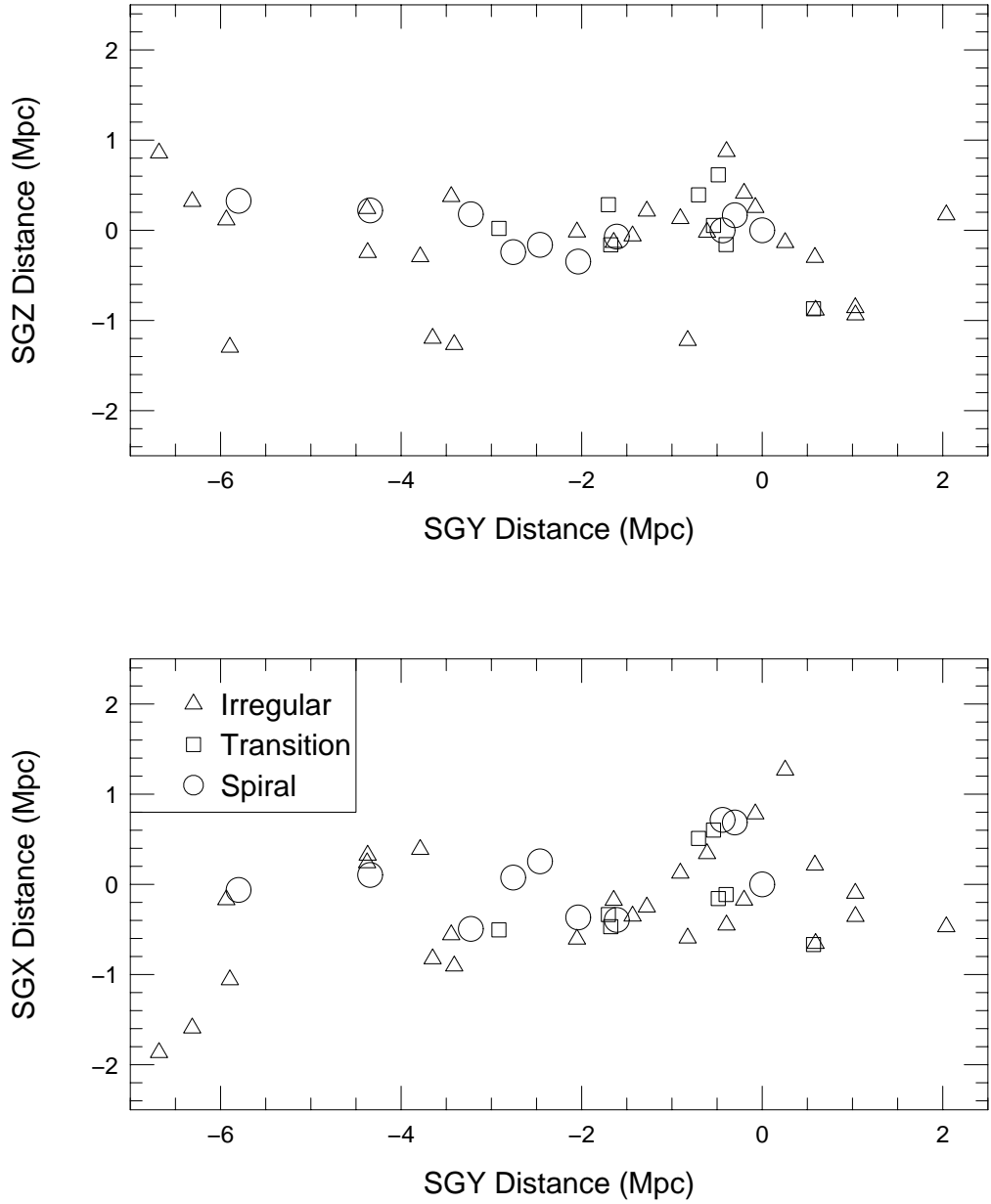}
\figcaption{
A comparison of the positions of the spirals (circles), dIs (triangles),
and transition type galaxies (squares) in the Local Group and Sculptor
Group plotted in super-galactic coordinates Z versus Y and X versus Y.
The Local Group and the Sculptor Group appear to join smoothly when
portrayed in this manner (see also Tully \& Fisher 1987 and Figure 9
of Jerjen et al.\ 1998).
\label{fig7}}
\end{figure}

\clearpage

\begin {figure}
\plotone{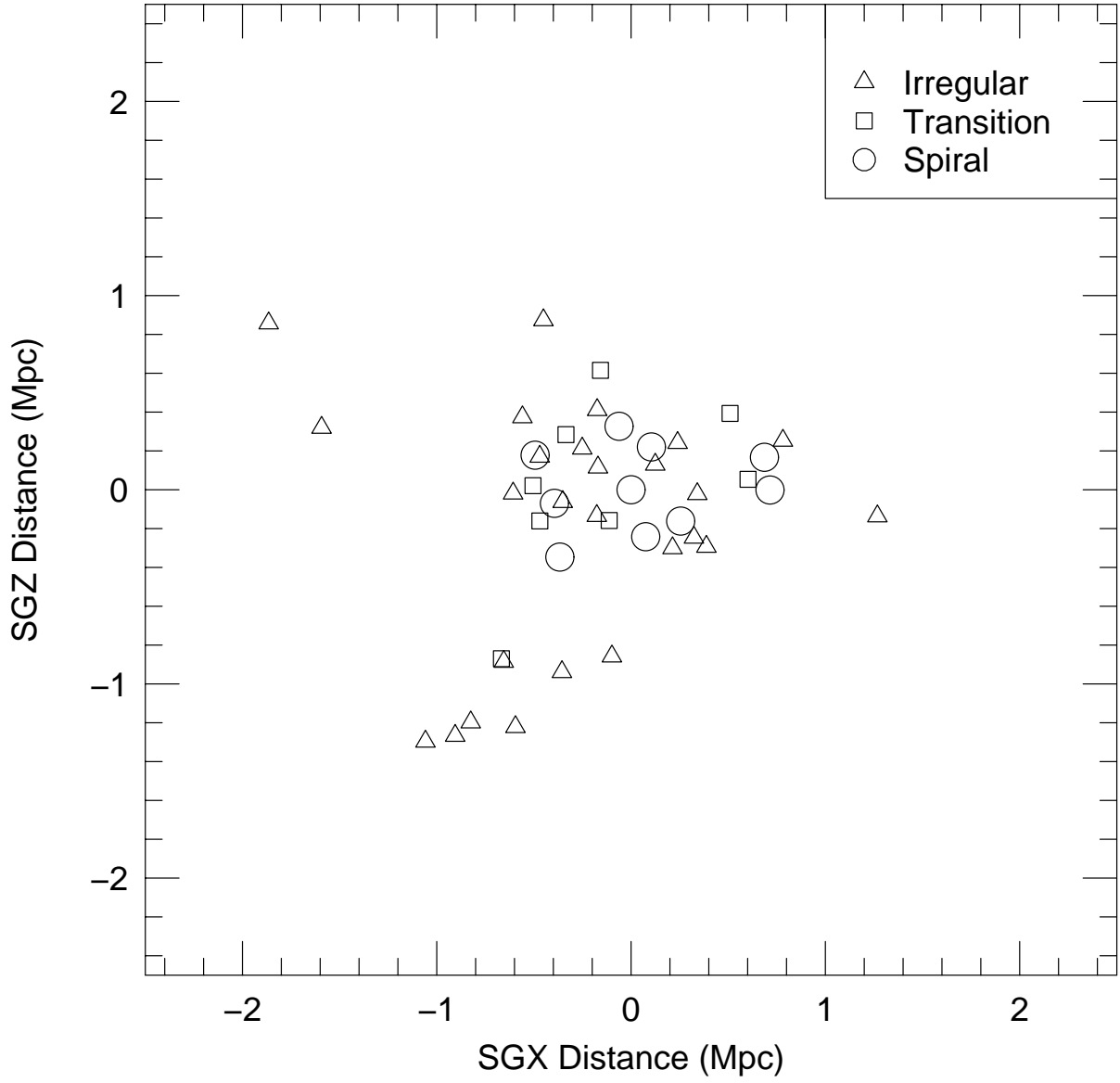}
\figcaption{
A comparison of the positions of the spirals (circles), dIs (triangles),
and transition type galaxies (squares) in the Local Group and Sculptor
Group plotted in super-galactic X and Z coordinates.  The dIs lie at
preferentially larger radii in this projection.
\label{fig8}}
\end{figure}

\clearpage

\begin {figure}
\plotone{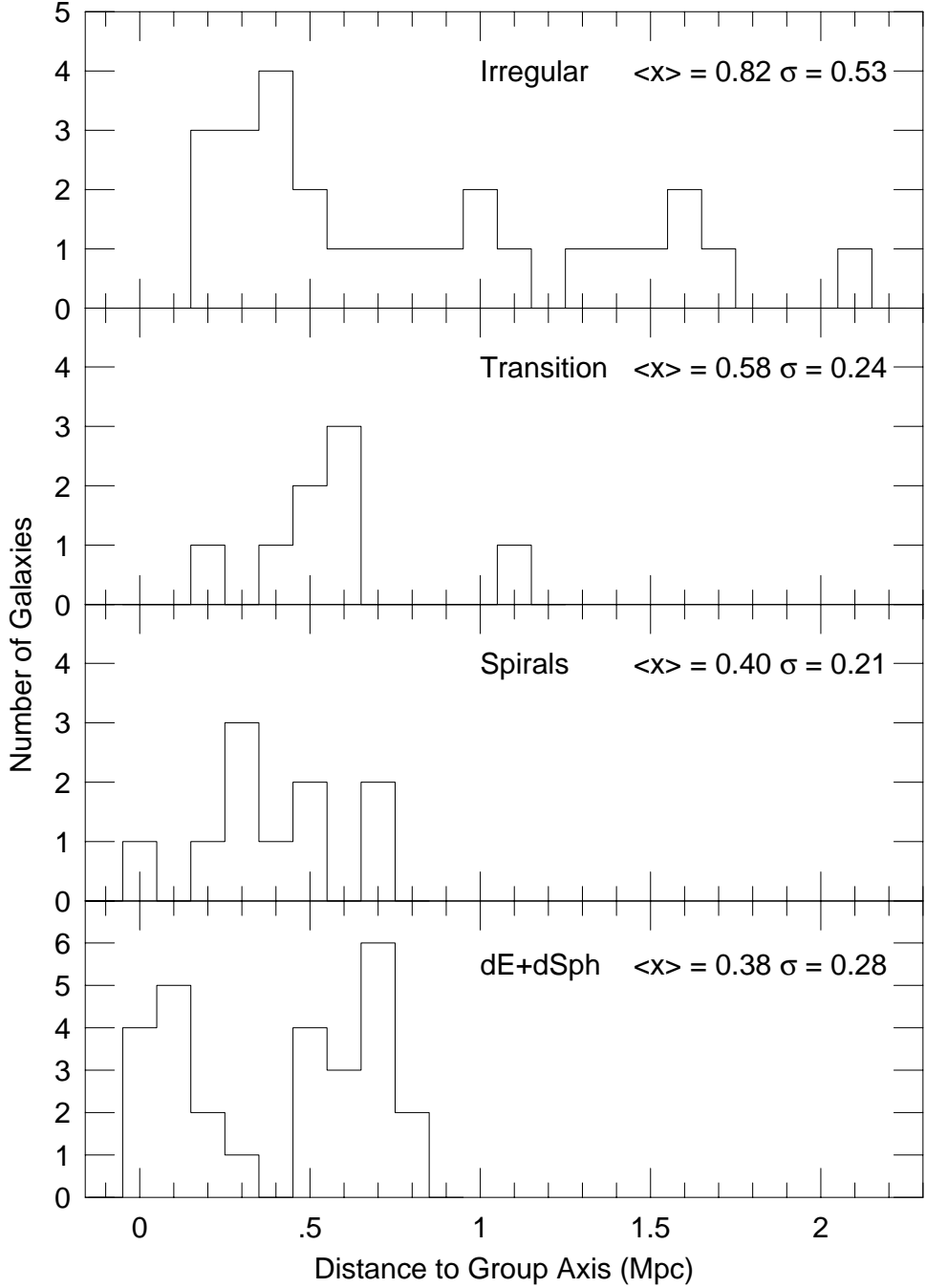}
\figcaption{
A histogram comparison of the radial distances of the dIs, transition,
spiral, and dE type galaxies in the Local Group and Sculptor
Group concentration (as described in the text and shown in Figure 8).
For each sample, the mean value and the standard deviation in the sample is given.
On average, the dIs lie at preferentially larger radii and the transition 
galaxy distances are comparable to those of the spiral and dE galaxies.
\label{fig9}}
\end{figure}

\clearpage

\begin {figure}
\plotone{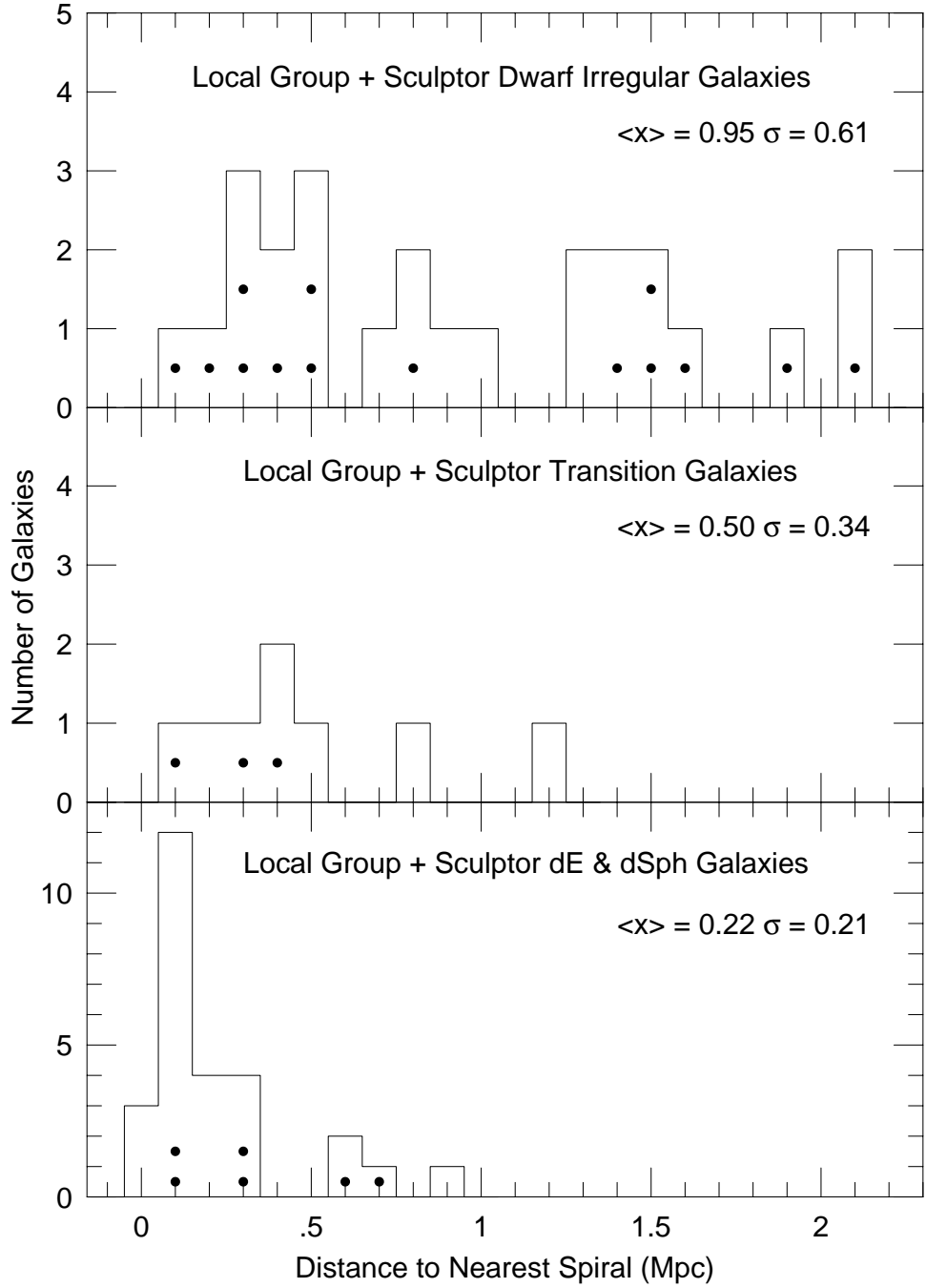}
\figcaption{
A histogram comparison of the distances to the nearest spiral galaxy
for the  dI, transition type, and dE galaxies in the Local Group and 
Sculptor Group.
For each sample, the mean value and the standard deviation in the sample is given.
On average, the dIs lie at preferentially larger distances from the 
spiral galaxies than the transition galaxies, and the transition galaxies
are at larger distances than the dEs.
The dots indicate the Sculptor Group members (see text). 
\label{fig10}}
\end{figure}

\end{document}